\documentclass[prd,amsmath,amssymb,superscriptaddress,floatfix,nofootinbib,10pt]{revtex4}
\usepackage{times}
\usepackage{amssymb,amsbsy,amsmath,amsfonts}
\usepackage{graphicx}
\usepackage{float}
\usepackage{color}
\usepackage{morefloats}
\usepackage{rotating}
\usepackage{srcltx}
\usepackage{slashed}
\usepackage{multirow}
\usepackage{verbatim}
\usepackage{hyperref}
\usepackage{tabularx}
\usepackage{bm}
\allowdisplaybreaks[4]

\usepackage{bbding}
\usepackage{threeparttable}

\newcommand{\PreserveBackslash}[1]{\let\temp=\\#1\let\\=\temp}
\newcolumntype{C}[1]{>{\PreserveBackslash\centering}p{#1}}
\newcolumntype{R}[1]{>{\PreserveBackslash\raggedleft}p{#1}}
\newcolumntype{L}[1]{>{\PreserveBackslash\raggedright}p{#1}}

\newcommand{\sj}[1]{{\color{black} #1}}

\begin{document}

\title{The role of meson interactions in the $D_s^{+}\rightarrow \pi^{+}\pi^{+}\pi^{-}\eta$ decay}

\author{Jing Song}
\email[E-mail me at: ]{song-jing@buaa.edu.cn}
\affiliation{School of Physics, Beihang University, Beijing, 102206, China}
\affiliation{Departamento de Física Teórica and IFIC, Centro Mixto Universidad de Valencia-CSIC Institutos de Investigación de Paterna,
Aptdo.22085, 46071 Valencia, Spain}

\author{ A. Feijoo}
\email[E-mail me at: ]{edfeijoo@ific.uv.es}
\affiliation{Departamento de Física Teórica and IFIC, Centro Mixto Universidad de Valencia-CSIC Institutos de Investigación de Paterna,
Aptdo.22085, 46071 Valencia, Spain}

\author{ E.Oset}
\email[E-mail me at: ]{oset@ific.uv.es}
\affiliation{Departamento de Física Teórica and IFIC, Centro Mixto Universidad de Valencia-CSIC Institutos de Investigación de Paterna,
Aptdo.22085, 46071 Valencia, Spain}

\begin{abstract}
We perform a theoretical study of the $D_s^{+}\to \pi^{+}\pi^{+}\pi^{-}\eta$ decay. We look first at the basic $D_s^{+}$ decay at the quark level from external and internal emission. Then hadronize a pair or two pairs of $q\bar{q}$ states to have mesons at the end. Posteriorly the pairs of mesons are allowed to undergo final state interaction, by means of which the $a_0(980)$, $f_0(980)$, $a_1(1260)$, and $b_1(1235)$ resonances are dynamically generated. The $G$-parity is used as a filter of the possible channels, and from those with negative $G$-parity only the ones that can lead to $\pi^{+}\pi^{+}\pi^{-}\eta$ at the final state are kept. Using transition amplitudes from the chiral unitary approach that generates these resonances, and a few free parameters, we obtain a fair reproduction of the six mass distributions reported in BESIII experiment.
\end{abstract}


\date{\today}

\maketitle
\section{Introduction}
$D$ meson decays into $3$ mesons have long being considered of good source of information to study the interaction of mesons~\cite{Boito:2009qd,Magalhaes:2011sh,Dedonder:2014xpa,Xie:2014tma,Magalhaes:2015fva,Sekihara:2015iha,Dias:2016gou,Sakai:2017iqs,Niecknig:2017ylb,Molina:2019udw,Hsiao:2019ait,Duan:2020vye,Roca:2020lyi,Toledo:2020zxj} (see review in~\cite{Oset:2016lyh}). The scattering mechanisms of final meson pairs are usually investigated in these works trying to obtain information on this interaction and resonances formed in the process. The $D$ decay into four mesons introduces a challenging task due to the additional meson pairs that require a consideration. In this paper we wish to do such a theoretical work on the $D_s^{+}\to \pi^{+}\pi^{+}\pi^{-}\eta$ reaction measured for the first time in~\cite{BESIII:2021aza} by the BESIII collaboration. One intriguing aspect of the analysis of~\cite{BESIII:2021aza} is the claim that the $D_s^{+}\to a^{+}_0(980)\rho^0$ decay mode proceeds via weak annihilation, with a rate about one order of magnitude bigger than ordinary weak-annihilation processes. In the order of the relevance of different weak decay mechanisms, weak-annihilation goes below external emission, internal emission and W-exchange~\cite{Chau:1982da,Chau:1987tk}. Hence, observing a reaction which proceeds via this mode with exceptionally large strength is certainly a relevant finding. Yet, there might be some ways to produce this mode indirectly, producing other intermediate states that lead to the desired final state through strong interaction transitions in the final states. This is one of the issues that we investigate here.

It is not the first time that such a thing happens, since in the study of $D_s^{+}\to \pi^{+}\pi^{0}\eta$ decay~\cite{BESIII:2019jjr}, the $\pi^{+}a_0(980)$ decay mode was also branded as an example of weak-annihilation with an abnormally large strength. Yet, in Ref.~\cite{Molina:2019udw} it was found that the process could be explained through a mechanism of internal emission, through the production of $K\bar{K}$ and the subsequent $K\bar{K}\to\pi\eta$ transition, dominated by the $a_0(980)$ resonance. An alternative explanation was provided in~\cite{Hsiao:2019ait} through a triangle mechanism where one has $D_s\to \rho^{+}\eta$, which proceed via external emission, followed by $\rho\to\pi\pi$ and fusion of $\pi\eta$ to give the $a_0(980)$ resonance. The same argumentation was followed in~\cite{Ling:2021qzl}, where the work of \cite{BESIII:2019jjr} was discussed and other possible mechanisms were considered. What is clear is that with either of the mechanisms of~\cite{Molina:2019udw,Hsiao:2019ait,Ling:2021qzl} one does not need weak-annihilation for the $D_s\to\pi a_0(980)$ production, and the strong interaction of the resulting mesons can lead to the desired final state. We will find a similar situation in the present reaction, where considering mechanisms of external and internal emission of different intermediate particles and allowing them to make transitions through final state interaction, we can obtain the desired final state. 

So far there is only one theoretical work which pays attention to this reaction~\cite{Yu:2021euw}. The work looks only to two particular decay channels, $D_s\to \rho^{0}a^+_0(980)\to \rho\pi\eta$ and $D_s\to \rho^{+}a^0_0(980)\to \rho^+K^+K^-$, using a triangle diagram like $D_s\to \pi\eta$ (virtual $\pi \to \pi\rho$), and fusion of $\pi\eta$ to give the $a_0(980)$ resonance. A similar mechanism with $K^+\bar{K}^0$ intermediate states in the loop instead of two pions is also considered. In both cases the primary products of the decay have a much smaller mass than the final ones, forcing them to be highly off shell. The work of~\cite{Yu:2021euw} finds reasonable results for the cases of $D_s\to \rho^{0}a^+\to \rho\pi\eta$ decay compared with the analysis of~\cite{BESIII:2021aza}, but with a rate for $D_s\to \rho^{+}a_0^0$ ($a_0\to K^+K^-$) one order of magnitude smaller. They hint at a possible misinterpretation of the data of~\cite{BESIII:2021aza} due to a possible contribution of $D_s\to f_0(980)\rho^{+}$.

Our aim in the present work is more ambitious, since we want to reproduce the six invariant mass distributions that have been reported in~\cite{BESIII:2021aza}, $M_{\pi^{+}\pi^{+}}$, $M_{\pi^{+}\pi^{-}}$, $M_{\pi^{+}\eta}$, $M_{\pi^{-}\eta}$, $M_{\pi^{+}\pi^{+}\pi^{-}}$, and $M_{\pi^{+}\pi^{-}\eta}$. The methodology is also different, we look at the reaction from the perspective that the different resonances that are observed in the analysis of the reaction, $f_0(500)$, $f_0(980)$, $a_0(980)$ and $a_1(1260)$, are obtained within the chiral unitary approach from the interaction of different mesons. In this sense, the scalar resonances $f_0(500)$, $f_0(980)$, and $a_0(980)$ are obtained from the interaction of pairs of pseudoscalar mesons in coupled channels~\cite{Oller:1997ti,Kaiser:1998fi,Locher:1997gr,Nieves:1999bx},  while the $a_1(1260)$ and other axial vector mesons are obtained from the interaction of the pairs of a pseudoscalar and a vector meson~\cite{Lutz:2003fm,Roca:2005nm,Geng:2006yb,Zhou:2014ila}. Then our procedure is as follows: we consider all possible decay mechanisms at the quark level and then proceed to create vectors and pseudoscalar recurring to the hadronization of $q\bar{q}$ pairs into a pair of mesons, pseudoscalar-pseudoscalar ($PP$) or vector-pseudoscalar ($VP$). After this, we allow the different $PP$ or $VP$ pairs to interact, leading to the $\pi^{+}\pi^{+}\pi^{-}\eta$ configuration finally. In the process of interaction different resonances are produced which are clearly visible in the different mass distributions. Our approach contains a few free parameters related to the strength of the different primary production processes, for which the order of magnitude is known, and a good reproduction of the six invariant mass distributions is obtained with considerably less freedom than in the partial wave analysis done in the experiment~\cite{BESIII:2021aza}. The relevant role played by the different resonances is then exposed.

\section{formalism}
\subsection{$G$-parity of the process}

The first realization in the $D_s^{+}\rightarrow \pi^{+}\pi^{+}\pi^{-}\eta$ reaction is that the $G$-parity of the final state is negative. Hence, after the weak interaction and prior to any final state interaction we must select only sates of $G$-parity negative. Pseudoscalar mesons and vector mesons without strangeness have given $G$-parity, $\eta$, $\rho$ positive, $\pi$, $\omega$, $\phi$ negative. $K$ or $K^*$ have no $G$-parity but pairs of them can have. The $G$-parity is defined as 
\begin{align}\label{Gparity}
G=Ce^{-i\pi I_2}; \, \, \, e^{-i\pi I_2}|I,I_3\rangle=(-1)^{I-I_3}|I,-I_3\rangle .
\end{align}
With our isospin doublet convention $(K^+, K^0)$, $(\bar{K}^0, -K^-)$ and $CK^+=K^-$, $CK^0=\bar{K}^0$, $(K^{*+}, K^{*0})$, $(\bar{K}^{*0}, -K^{*-})$ and $CK^{*+}=-K^{*-}$, $CK^{*0}=-\bar{K}^{*0}$, we have the $G$-parity acting over the $K$, $K^*$ states as shown in Table~\ref{tableI}. 

\begin{table}[H]
\centering
 \caption{$G$-parity acting on $K$ and $K^*$ states.}\label{tableI}
\setlength{\tabcolsep}{14pt}
\begin{tabular}{cccccccccc}
\hline
\hline
~~& $K^+$ & $K^0$ & $\bar{K}^0$ & $K^-$ & ~~ & $K^{*+}$ & $K^{*0}$ & $\bar{K}^{*0}$ & $K^{*-} $ \\
\hline
$G(K_i)$& $\bar{K}^0$ & $-K^-$ & $-K^+$ & $K^0$ & ~~ & $-\bar{K}^{*0}$ & $K^{*-}$ & $K^{*+}$ & $-K^{*0}$  \\
\hline
\hline
\end{tabular}
\end{table}

\subsection{External emission with one hadronization}
\begin{figure}[H]
  \centering
  \includegraphics[width=0.3\textwidth]{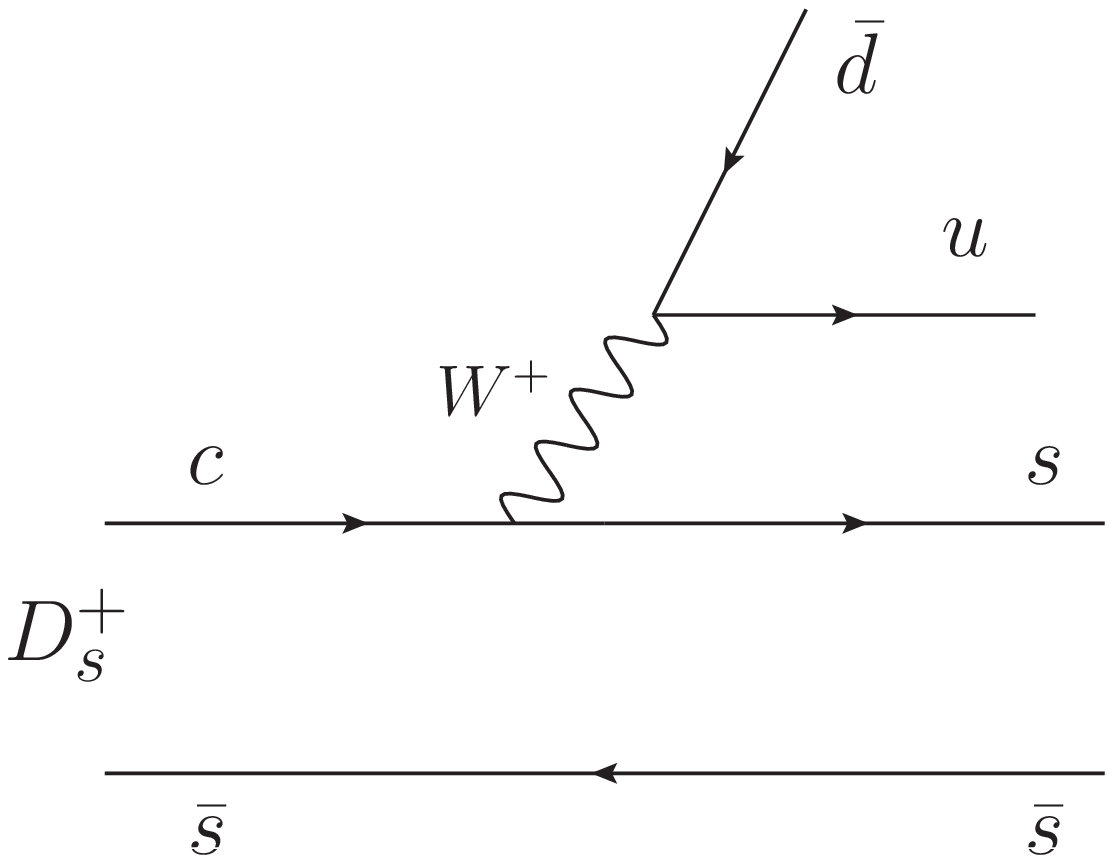}
  \includegraphics[width=0.28\textwidth]{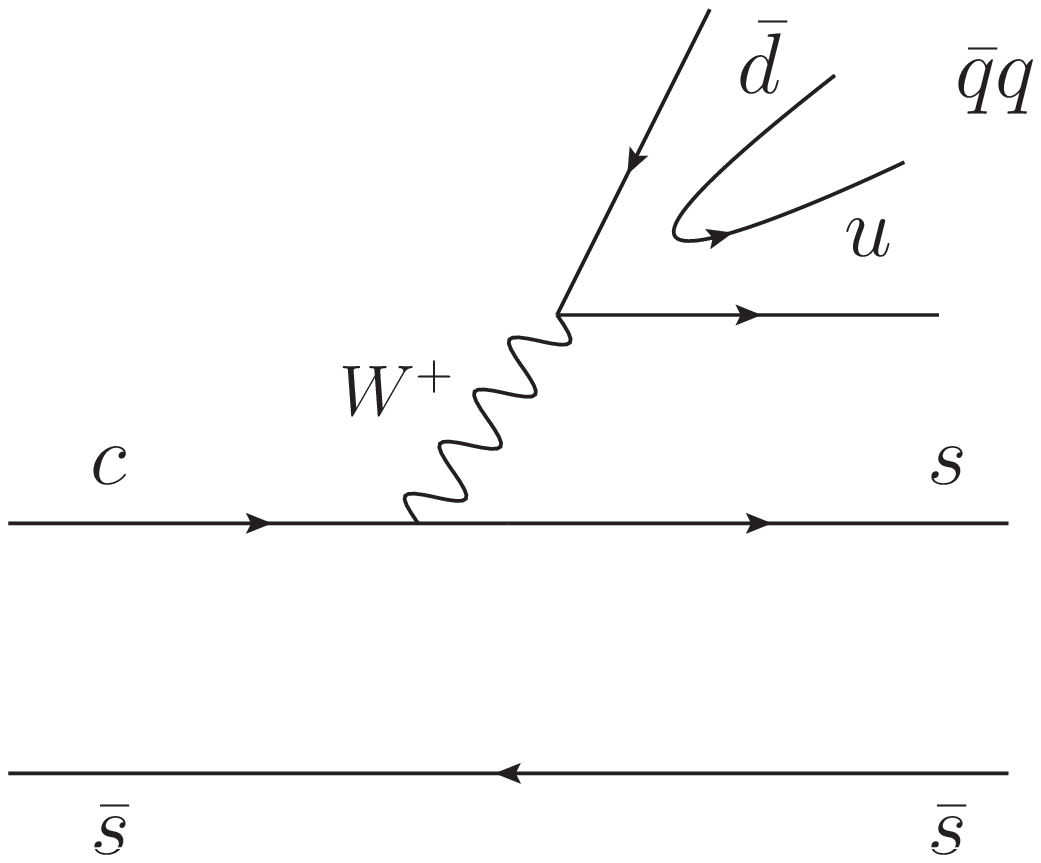}
  \includegraphics[width=0.3\textwidth]{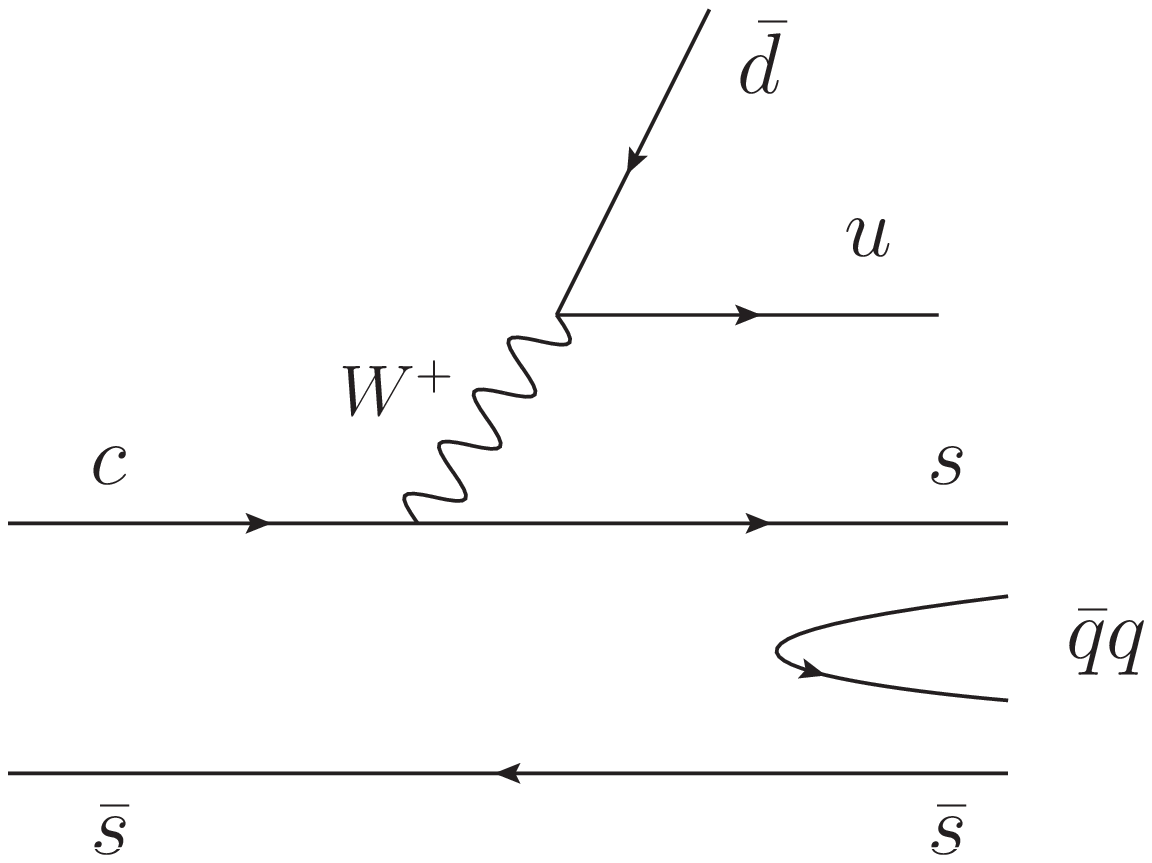}
  \caption{External emission mechanism for $D_s^{+}$ decay at the quark level. (a) basic mechanism; (b) hadronization of the $u\bar{d}$ component; (b) hadronization of the $s\bar{s}$ component.}\label{fig1}
\end{figure}

We show in Fig.~\ref{fig1} the Cabibbo favored process of external emission at the quark level.
 Since we need four particles in the final states we need to produce a vector meson, which will decay to two pseudoscalars, and a pair of pseudoscalars. We have several options:
 
  \begin{enumerate}
  \item Hadronize $\bar{d}u$ with $PP$ and $s\bar{s}$ giving a vector;
  \item Produce a vector from $\bar{d}u$   and hadronize $s\bar{s}$ to two pseudoscalars;
  \item  Hadronize $\bar{d}u$ with $VP$ or $PV$ and $s\bar{s}$ to two pseudoscalars;
  \item Hadronize $\bar{s}s$ to $VP$ or $PV$ and $\bar{d}u$ a pseudoscalar.
\end{enumerate} 

Let us see the consequences. For this we would need the representation of $q\bar{q}$ in terms of mesons, which is given by 

\begin{align}\label{meson}
\centering
P=
\left(
  \begin{array}{ccc}
    \frac{\pi^0}{\sqrt{2}}+\frac{\eta}{\sqrt{3}}+\frac{\eta'}{\sqrt{6}} & \pi^{+} & K^{+} \\
    \pi^{-} & \frac{-\pi^0}{\sqrt{2}}+\frac{\eta}{\sqrt{3}}+\frac{\eta'}{\sqrt{6}} & K^0 \\
    K^{-} & \bar{K}^0 & -\frac{\eta}{\sqrt{3}}+\frac{2\eta'}{\sqrt{3}} \\
  \end{array}
\right),~~~~
V=
\left(
  \begin{array}{ccc}
    \frac{\rho^0}{\sqrt{2}}+\frac{\omega}{\sqrt{2}} & \rho^{+} & K^{*+} \\
    \rho^{-} & \frac{-\rho^0}{\sqrt{2}}+\frac{\omega}{\sqrt{2}} & K^{*0} \\
    K^{*-} & \bar{K}^{*0} & \phi \\
  \end{array}
\right),
\end{align}
where we have taken the ordinary $\eta-\eta'$ mixing of Ref.~\cite{Bramon:1992kr}.

Let us see what one obtains with the four options above. (We neglect $\eta'$ which plays no role in these processes.)\\

1)\begin{align}
    u\bar{d} \to \sum_{i} u\bar{q}_iq_i\bar{d}=&\sum_{i}P_{1i}P_{i2}=(P^2)_{12}=(\frac{\pi^{0}}{\sqrt{2}}+\frac{\eta}{\sqrt{3}})\pi^{+}+\pi^{+}(\frac{-\pi^{0}}{\sqrt{2}}+\frac{\eta}{\sqrt{3}})+K^{+}\bar{K}^{0}
\end{align}
On the other hand the vector for $s\bar{s}$ is $\phi$ which decays to $K\bar{K}$ but not to $\pi^{+}\pi^{-}$ (different $G$-parity). Hence, this process cannot lead to our final state of $\pi^{+}\pi^{+}\pi^{-}\eta$.\\

2)
\begin{align}\label{8_1}
    s\bar{s} \to \sum_{i} s\bar{q}_iq_i\bar{s}=&\sum_{i}P_{3i}P_{i3}=(P^2)_{33}=K^{-}K^{+}+\bar{K}^{0}K^{0}+\frac{\eta\eta}{3}
\end{align}
and $\bar{d}u$ as a vector is $\rho^+$. The vector $\rho^+$ can give  $\pi^+\pi^0$ but $s\bar{s}$ has $I=0$ and hadronization does not change the isospin, which means that the combination of Eq.~(\ref{8_1}) cannot give $\pi^0\eta$. Once again, this mechanism cannot produce our final state.\\

3a) We hadronize $u\bar{d}$ with $VP$ and have 
\begin{align}\label{8_2}
    u\bar{d} \to \sum_{i} u\bar{q}_iq_i\bar{d}=&\sum_{i}V_{1i}P_{i2}=(VP)_{12}=(\frac{\rho^{0}}{\sqrt{2}}+\frac{\omega}{\sqrt{2}})\pi^{+}+\rho^{+}(\frac{-\pi^{0}}{\sqrt{2}}+\frac{\eta}{\sqrt{3}})+K^{*+}\bar{K}^0    
\end{align}
and $s\bar{s}$ will give rise to the $\eta$. According to the $\eta-\eta'$ mixing of Ref.~\cite{Bramon:1992kr} one has
\begin{align}\label{8_3}
  s\bar{s}=-\frac{1}{\sqrt{3}}\eta+\sqrt{\frac{2}{3}}\eta'
\end{align}
We can see that there are already candidates, since $\rho^0\to\pi^{+}\pi^{-}$ and we can have $\pi^{+}\pi^{+}\pi^{-}\eta$.\\

3b) We hadronize $u\bar{d}$ with $PV$ and have 
\begin{align}\label{8_4}
    u\bar{d} \to \sum_{i} u\bar{q}_iq_i\bar{d}=&\sum_{i}P_{1i}V_{i2}=(PV)_{12}=(\frac{\pi^{0}}{\sqrt{2}}+\frac{\eta}{\sqrt{3}})\rho^{+}+\pi^{+}(\frac{-\rho^{0}}{\sqrt{2}}+\frac{\omega}{\sqrt{2}})+K^{+}\bar{K}^{*0}    
\end{align}
and once again   with  $\rho^0\to\pi^{+}\pi^{-}$, the extra $\pi^{+}$ and $\eta$ from $s\bar{s}$, the combination  can give the desired final state.

Now, an inspection of Eqs.(\ref{8_2}), (\ref{8_4}) shows immediately that these states mix the $G$-parity, which is not surprising since they come from a weak process that does not conserve isospin. However, we can make good $G$-parity states by means of the linear combinations $(VP\pm PV)$. Indeed,
\begin{align}\label{9_1}
    (VP)_{12}-(PV)_{12}=&\sqrt{2}\rho^{0}\pi^{+}-\sqrt{2}\rho^{+}\pi^{0}+K^{*+}\bar{K}^0-K^{+}\bar{K}^{*0}
\end{align}
\begin{align}\label{9_2}
    (VP)_{12}+(PV)_{12}=&\sqrt{2}\omega\pi^{+}+\frac{2}{\sqrt{3}}\rho^{+}\eta+K^{*+}\bar{K}^0+K^{+}\bar{K}^{*0}    
\end{align}
If we look at Table~\ref{tableI} we can see that $(VP)_{12}-(PV)_{12}$ has negative $G$-parity, while $VP+PV$ has positive $G$-parity. Hence, it is the $(VP-PV)_{12}$ combination of Eq.~(\ref{9_1}) the one we shall take to produce the final state $\pi^{+}\pi^{+}\pi^{-}\eta$. Thus we consider the state
\begin{align}\label{9_3}
    |HE3'\rangle= (VP-PV)_{12}=-\frac{1}{\sqrt{3}}\eta\Big(\sqrt{2}\rho^{0}\pi^{+}-\sqrt{2}\rho^{+}\pi^{0}+K^{*+}\bar{K}^0-K^{+}\bar{K}^{*0}\Big)
\end{align}
but the $\rho^{+}\pi^{0}\eta$ combination going to $\pi^{+}\pi^{0}\pi^{0}\eta$ will not contribute. We can therefore take 
\begin{align}\label{new_Eq}
    |HE3\rangle=-\sqrt{\frac{2}{3}}\eta\rho^{0}\pi^{+}-\frac{1}{\sqrt{3}}\eta(K^{*+}\bar{K}^0-K^{+}\bar{K}^{*0})
\end{align}

4a)
\begin{align}\label{9_4}
    s\bar{s} \to \sum_{i} s\bar{q}_iq_i\bar{s}=&\sum_{i}V_{3i}P_{i3}=(VP)_{33}=K^{*-}K^{+}+\bar{K}^{*0}K^{0}-\phi\frac{\eta}{\sqrt{3}}
\end{align}
and $\bar{d}u$ will be a $\pi^{+}$. By looking again to Table~\ref{tableI} we can see that the former combination together  with $\pi^{+}$ has not a well defined $G$-parity.\\

4b)
\begin{align}\label{10_1}
    s\bar{s} \to \sum_{i} s\bar{q}_iq_i\bar{s}=&\sum_{i}P_{3i}V_{i3}=(PV)_{33}=K^{-}K^{*+}+\bar{K}^{0}K^{*0}-\phi\frac{\eta}{\sqrt{3}}
\end{align}
which , again, has no defined $G$-parity.

We construct the $VP\pm PV$ combinations and find  
\begin{align}\label{10_2}
    (VP)_{33}-(PV)_{33}=&K^{*-}K^{+}-K^{-}K^{*+}+\bar{K}^{*0}K^{0}-\bar{K}^{0}K^{*0}
\end{align}
\begin{align}\label{10_3}
    (VP)_{33}+(PV)_{33}=&K^{*-}K^{+}+K^{-}K^{*+}+\bar{K}^{*0}K^{0}+\bar{K}^{0}K^{*0}-\frac{2}{\sqrt{3}}\phi\eta
\end{align}
Once again we see that the $VP-PV$ combination together with $\pi^{+}$ has $G$-parity negative, while $VP+PV$ and $\pi^{+}$ has $G$-parity positive. We can think of making the transition of $(VP-PV)_{33}$ to $\rho^0\eta$ to complete $\rho^{0}\eta\pi^{+}\to\pi^{+}\pi^{-}\eta\pi^{+}$. However, this is not possible since the combination $VP-PV$, coming from $s\bar{s}$, has $I=0$ and hence cannot go to $\rho\eta$. Indeed, the $I^G=0^+$ resonance coming from $VP-PV$ combinations is the $f_1(1285)$(see Table~\ref{tableII}), which only couples to $K^{*}\bar{K}$, $\bar{K}^{*}K$ channels~\cite{Roca:2005nm,Zhou:2014ila}. Thus, from all possible states  coming from hadronization with external emission, only one, $|HE3\rangle$ of Eq.~(\ref{new_Eq}) can lead to our desired final state.
\begin{table}[H]
\centering
 \caption{non strange axial vector resonances generated by the $VP$, $PV$ interaction~\cite{Roca:2005nm,Lutz:2003fm}.}\label{tableII}
\setlength{\tabcolsep}{14pt}
\begin{tabular}{cccccc}
\hline
\hline
~~& $h_1(1173)$ & $h_1(1380)$ & $b_1(1235)$ & $a_1(1260)$ & $f_1(1285)$ \\
\hline
$I^G$ & $0^-$ & $0^-$ & $1^+$ & $1^-$ & $0^+$  \\
\hline
\hline
\end{tabular}
\end{table}

\subsection{Internal emission with one hadronization}
\begin{figure}[H]
  \centering
  \includegraphics[width=0.3\textwidth]{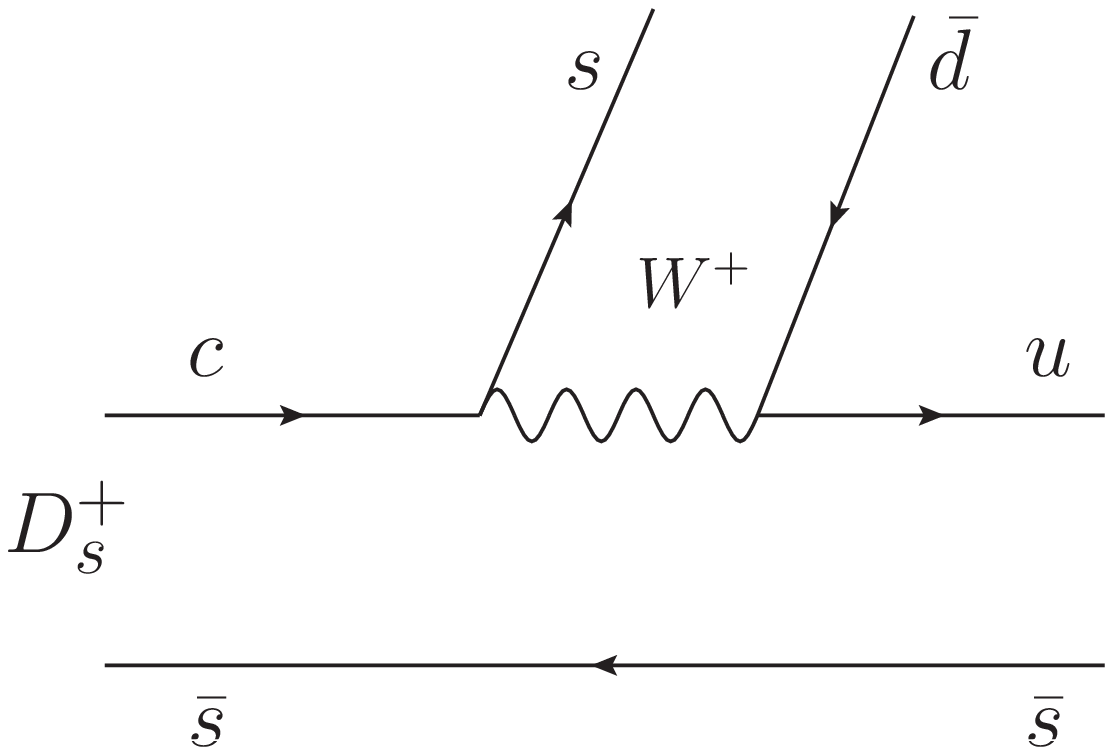}
  \includegraphics[width=0.28\textwidth]{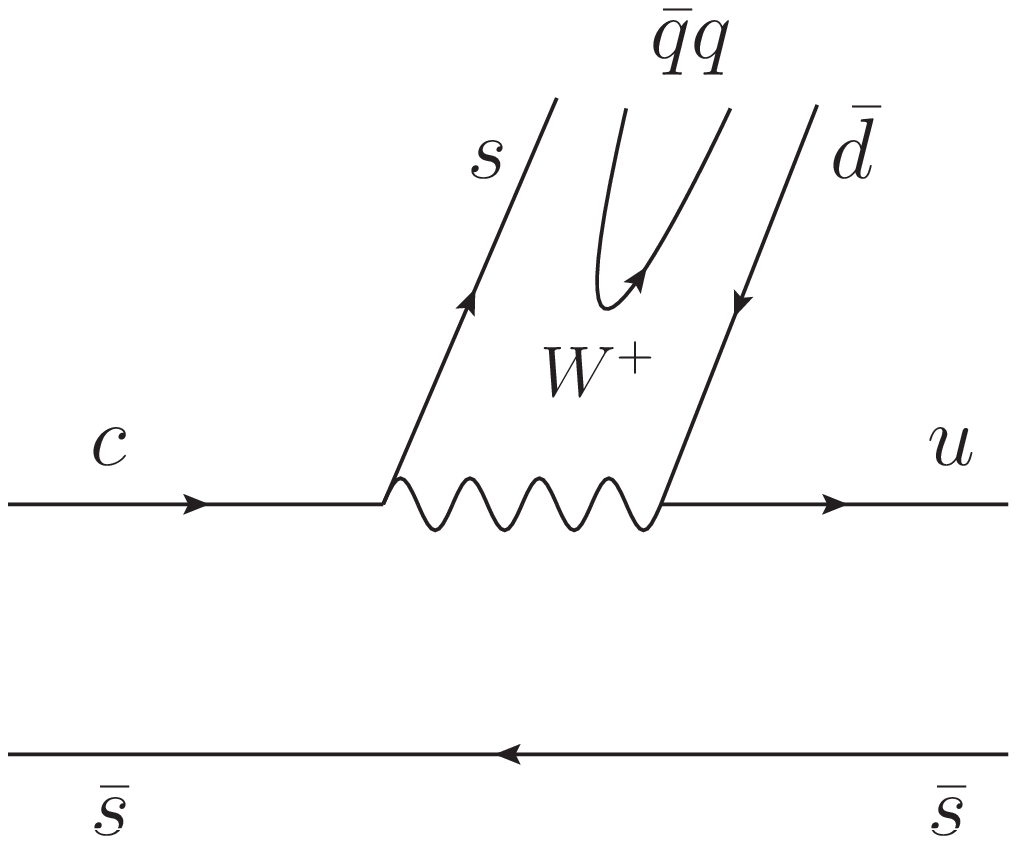}
  \includegraphics[width=0.3\textwidth]{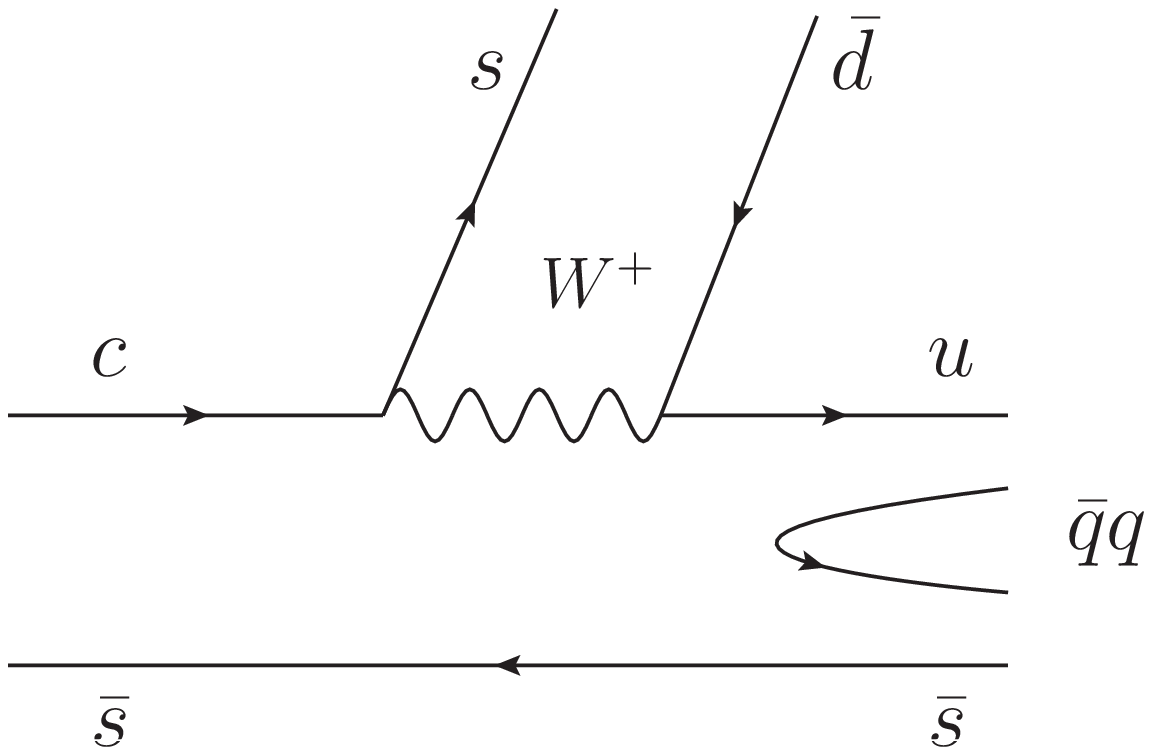}
  \caption{Internal emission mechanism for $D_s^{+}$ decay at the quark level. (a) basic mechanism; (b) hadronization of the $s\bar{d}$ component; (b) hadronization of the $u\bar{s}$ component.}\label{fig2}
\end{figure}

We look now at the process of Fig.~\ref{fig2} for the $D_s^+$ decay,
 We follow now the same strategy as before:
 \begin{enumerate}
  \item Hadronize $s\bar{d}$ with $PP$ and $u\bar{s}$ is a vector;
  \item Hadronize $u\bar{s}$ to $PP$ and $s\bar{d}$ is a vector;
  \item  Hadronize $s\bar{d}$ to $VP$, $PV$ and $u\bar{s}$ is a pseudoscalar;
  \item Hadronize $u\bar{s}$ to $VP$, $PV$ and $s\bar{d}$ is a pseudoscalar.
\end{enumerate} 

Let us see these possibilities in detail\\

1)
\begin{align}\label{11_1}
    s\bar{d} \to \sum_{i} s\bar{q}_iq_i\bar{d}=&\sum_{i}P_{3i}P_{i2}=(P^2)_{32}\\\nonumber
    =&K^{-}\pi^{+}+(-\frac{\pi^{0}}{\sqrt{2}}+\frac{\eta}{\sqrt{3}})\bar{K}^{0}-\frac{\eta}{\sqrt{3}}\bar{K}^{0}=K^{-}\pi^{+}-\frac{\pi^{0}}{\sqrt{2}}\bar{K}^{0}
\end{align}
and the $u\bar{s}$ component gives ${K}^{*+}$.\\

2)
\begin{align}\label{11_2}
    u\bar{s} \to \sum_{i} u\bar{q}_iq_i\bar{s}=&\sum_{i}P_{1i}P_{i3}=(P^2)_{13}\\\nonumber
    =&(\frac{\pi^{0}}{\sqrt{2}}+\frac{\eta}{\sqrt{3}})K^{+}+\pi^{+}K^{0}-\frac{\eta}{\sqrt{3}}K^{+}=\frac{\pi^{0}}{\sqrt{2}}K^{+}+\pi^{+}K^{0}
\end{align}
and the $s\bar{d}$ component gives $\bar{K}^{*0}$. Neither $(P^2)_{32}K^{*+}$ nor $(P^2)_{13}\bar{K}^{*0}$  have good $G$-parity, but we make again the combinations
\begin{align}\label{12_1}
    (P^2)_{32}{K}^{*+}-(P^2)_{13}\bar{K}^{*0}
    =&(K^{-}\pi^{+}-\bar{K}^{0}\frac{\pi^{0}}{\sqrt{2}})K^{*+}-(\frac{\pi^{0}}{\sqrt{2}}K^{+}+\pi^{+}K^{0})\bar{K}^{*0}\\\nonumber
    =&\pi^{+}(K^{*+}K^{-}-\bar{K}^{*0}K^{0})-\frac{\pi^{0}}{\sqrt{2}}(K^{*+}\bar{K}^{0}+\bar{K}^{*0}K^{+})
    \end{align}
\begin{align}\label{12_2}
    (P^2)_{32}{K}^{*+}+(P^2)_{13}\bar{K}^{*0}
    =&(K^{-}\pi^{+}-\bar{K}^{0}\frac{\pi^{0}}{\sqrt{2}})K^{*+}+(\frac{\pi^{0}}{\sqrt{2}}K^{+}+\pi^{+}K^{0})\bar{K}^{*0}\\\nonumber
    =&\pi^{+}(K^{*+}K^{-}+\bar{K}^{*0}K^{0})+\frac{\pi^{0}}{\sqrt{2}}(\bar{K}^{*0}K^{+}-K^{*+}\bar{K}^{0}).
    \end{align}
    
By looking at Table~\ref{tableI}, we see that the ($-$) combination has $G$-parity negative, while the ($+$)  combination has $G$-parity positive. Note also that since we have produced the quarks $s\bar{s}\bar{d}u$ in Fig.~\ref{fig2}, this has $|I,I_3\rangle=|1,1\rangle$ for all the combinations. In Eq.~(\ref{12_1}) which has $G$-parity negative, $K^{*+}K^{-}-\bar{K}^{*0}K^{0}$ has $G$-parity positive. On the other hand, this combination has $I=1,I_3=0$, hence, according to Table~\ref{tableII}, this combination is the one that creates the $b_1(1235)$, and thus Eq.~(\ref{12_1}) gives us a good combination. Since we do not want a $\pi^{0}$, at the end we choose the $\pi^{+}$ ($K^{*+}K^{-}-\bar{K}^{*0}K^{0}$) combination, corresponding to $\pi^{+}b_1^0$, then the $b_1^0$ decays to $\rho^0\eta$ (see Table~VII of Ref.~\cite{Roca:2005nm}), $\rho^0\to\pi^{+}\pi^{-}$ and we have the desired final state.

We then select the hadronic component
\begin{align}\label{12_3}
       |HI12\rangle= \pi^{+}(K^{*+}K^{-}-\bar{K}^{*0}K^{0})
\end{align}\\

3a) $VP$:
\begin{align}\label{13_1}
    s\bar{d} \to \sum_{i} s\bar{q}_iq_i\bar{d}=&\sum_{i}V_{3i}P_{i2}=(VP)_{32}=K^{*-}\pi^{+}+(-\frac{\pi^{0}}{\sqrt{2}}+\frac{\eta}{\sqrt{3}})\bar{K}^{*0}+\phi\bar{K}^{0}
\end{align}
together with $K^{+}$.\\

3b) $PV$:
\begin{align}\label{13_2}
    s\bar{d} \to \sum_{i} s\bar{q}_iq_i\bar{d}=&\sum_{i}P_{3i}V_{i2}=(PV)_{32}=K^{-}\rho^{+}+(-\frac{\rho^{0}}{\sqrt{2}}+\frac{\omega}{\sqrt{2}})\bar{K}^{0}-\bar{K}^{*0}\frac{\eta}{\sqrt{3}}
\end{align}
together with $K^{+}$.

Unlike we had before, we do not make good combinations now for $G$-parity positive or negative. This is done when we consider the 4a) and 4b) combinations below.\\

4a) $VP$:
\begin{align}
    u\bar{s} \to \sum_{i} u\bar{q}_iq_i\bar{s}=&\sum_{i}V_{1i}P_{i3}=(VP)_{13}=(\frac{\rho^{0}}{\sqrt{2}}+\frac{w}{\sqrt{2}})K^{+}+\rho^{+}K^{0}-{K}^{*+}\frac{\eta}{\sqrt{3}}
\end{align}
together with $\bar{K}^{0}$.\\

4b) $PV$:
\begin{align}
    u\bar{s} \to \sum_{i} u\bar{q}_iq_i\bar{s}=&\sum_{i}P_{1i}V_{i3}=(PV)_{13}=(\frac{\pi^{0}}{\sqrt{2}}+\frac{\eta}{\sqrt{3}})K^{*+}+\pi^{+}K^{*0}+K^{+}\phi
\end{align}
together with $\bar{K}^{0}$.

We find now four combinations from 3a), 3b) and 4a) 4b), two with positive $G$-parity and two with negative $G$-parity. Those of negative $G$-parity are:
\begin{align}\label{14_1}
    &K^{+}(VP)_{32}-\bar{K}^{0}(PV)_{13}=\pi^{+}(K^{*-}K^{+}-K^{*0}\bar{K}^{0})-\frac{\pi^{0}}{\sqrt{2}}(\bar{K}^{*0}K^{+}+K^{*+}\bar{K}^{0})+\frac{\eta}{\sqrt{3}}(\bar{K}^{*0}K^{+}-K^{*+}\bar{K}^{0})
\end{align}
\begin{align}\label{14_2}
    &\bar{K}^{0}(VP)_{13}-K^{+}(PV)_{32}=\rho^{+}(K^{0}\bar{K}^{0}-K^{+}K^{-})+{\sqrt{2}}{\rho^{0}}(K^{+}\bar{K}^{0})-\frac{\eta}{\sqrt{3}}(\bar{K}^{0}K^{*+}-\bar{K}^{*0}K^{+})
\end{align}
and in Eq.~(\ref{14_1}) the $\pi^0$ term will not contribute. Similarly, the $K^{0}\bar{K}^{0}-K^{+}K^{-}$ in Eq.~(\ref{14_2}) has $I=1$ and corresponds to the $a_0$ with zero charge that decays to $\pi^0\eta$. Together with $\rho^+$, we would have $\pi^{+}\pi^{0}\pi^{0}\eta$ which is not the desired final state. Hence we have two good combinations.
\begin{align}\label{14_3}
   |HI3213\rangle =\pi^{+}(K^{*-}K^{+}-K^{*0}\bar{K}^{0})+\frac{\eta}{\sqrt{3}}(\bar{K}^{*0}K^{+}-K^{*+}\bar{K}^{0})
\end{align}

\begin{align}\label{14_4}
   |HI1332\rangle = {\sqrt{2}}{\rho^{0}}K^{+}\bar{K}^{0}-\frac{\eta}{\sqrt{3}}(\bar{K}^{0}K^{*+}-\bar{K}^{*0}K^{+})
\end{align}

Let us inspect these terms. The combination $K^{*-}{K}^{+}-K^{*0}\bar{K}^{0}$ accompanying $\pi^+$ in Eq.~(\ref{14_1}) has $G$-parity positive and is a mixture of $I=0,1$. According to Table~\ref{tableII} it could contribute to produce the $b_1(1235)$ or the $f_1(1285)$, but only the $b_1$ decays to $\rho\eta$, hence we must project that state over the $b_1$. On the other hand, the combination $\bar{K}^{*0}K^{+}-K^{*+}\bar{K}^{0}$ has $I=1$, $I_3=1$, and with negative $G$-parity it corresponds to the $a_1(1260)$. Finally, the  $K^{+}\bar{K}^{0}$ accompanying $\rho^0$ in Eq.~(\ref{14_4}) has $I=1$, $I_3=1$ and negative $G$-parity and corresponds to the $a_0(980)$ resonance. As we can see, we have obtained terms that lead us to the $\pi^{+}\pi^{+}\pi^{-}\eta$ final state through the excitation of the $b_1(1235)$, $a_1(1260)$, and $a_0(980)$ resonances, which are well seen in the mass spectra of the experiment~\cite{BESIII:2021aza}.

We have obtained $4$ suitable states $|HE3\rangle$, $|HI12\rangle$, $|HI3213\rangle$, and $|HI1332\rangle$. We shall give weights $1$ to $|HE3\rangle$, $\alpha$ to $|HI3213\rangle$, $\beta$ to $|HI1332\rangle$, and $\gamma$ to $|HI12\rangle$, up to a global normalization factor $C$, and we get the contribution from one hadronization:
\begin{align}\label{15_1}
   |H1\rangle\equiv  C\bigg[&-\sqrt{\frac{2}{3}}\eta\rho^{0}\pi^{+}+ \frac{\eta}{\sqrt{3}}(1+\alpha+\beta)(\bar{K}^{*0}K^{+}-K^{*+}\bar{K}^{0})\\\nonumber
        &+ \sqrt{2}\beta \rho^{0}K^{+}\bar{K}^{0}
        + \alpha \pi^{+}(K^{*-}K^{+}-K^{*0}\bar{K}^{0})
        - \gamma \pi^{+}(\bar{K}^{*0}K^{0}-K^{*+}K^{-})
        \bigg].
\end{align}
The first term in the former equation is a tree level contribution, then $\rho^0\to\pi^{+}\pi^{-}$ and we shall have $\pi^{+}\pi^{+}\pi^{-}\eta$, the desired final state. The combinations with $K$, $K^{*}$ will make transitions to other states to complete the $\pi^{+}\pi^{+}\pi^{-}\eta$ final state and we assume these transitions to be dominated by the corresponding resonances that they form within the chiral unitary approach.

\begin{figure}[H]
  \centering
  \includegraphics[width=0.25\textwidth]{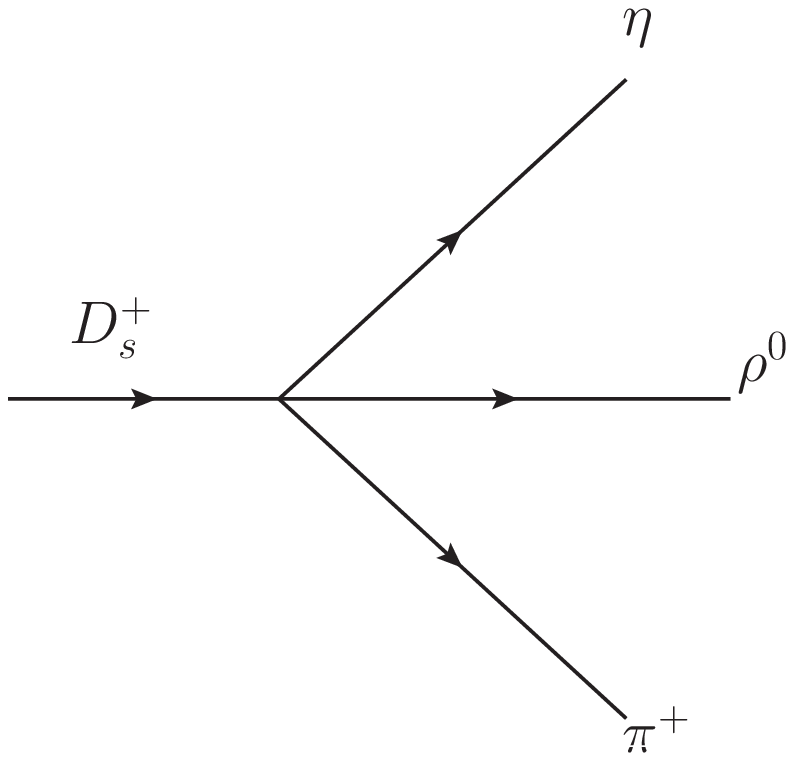}\qquad\qquad\qquad\qquad
  \includegraphics[width=0.4\textwidth]{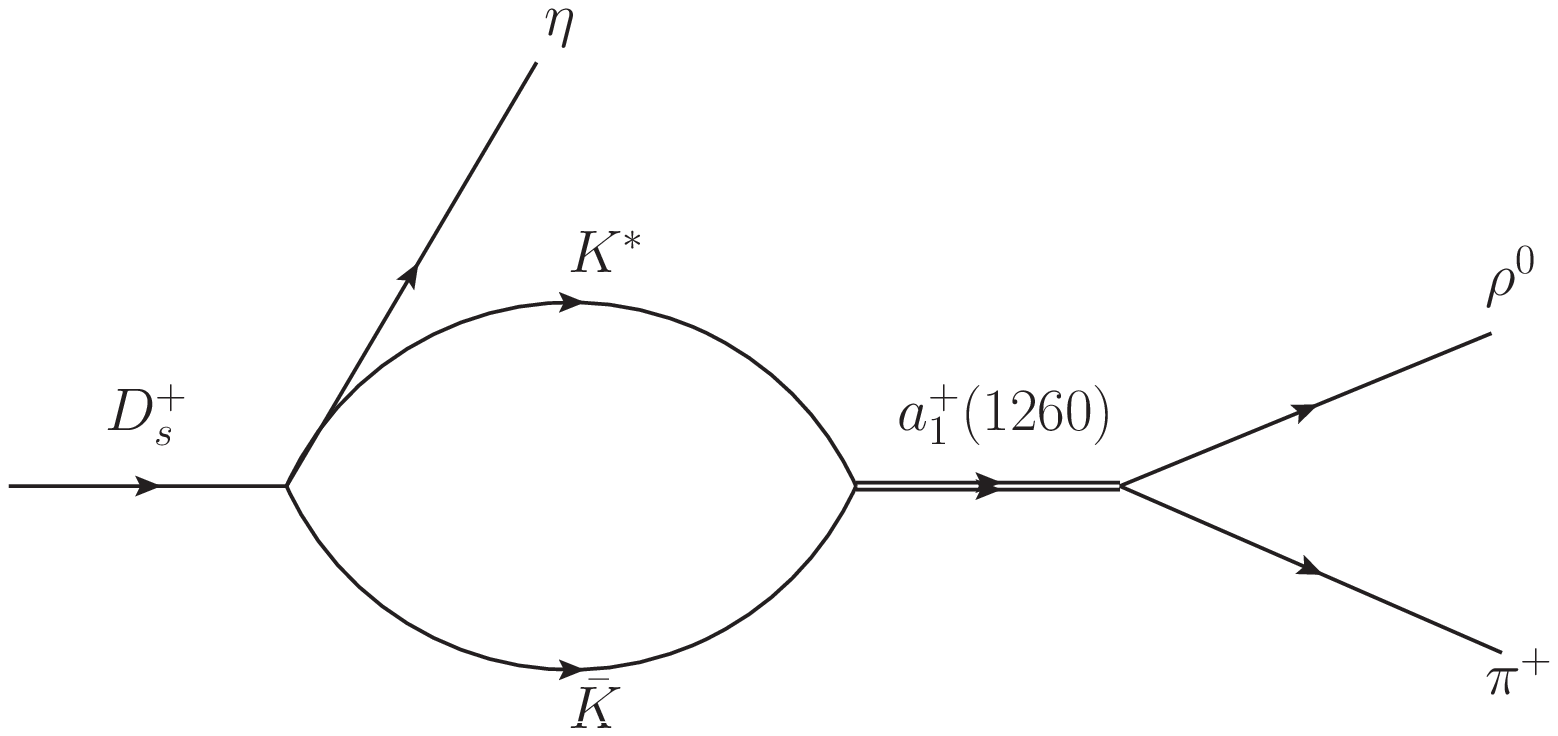}
  \includegraphics[width=0.4\textwidth]{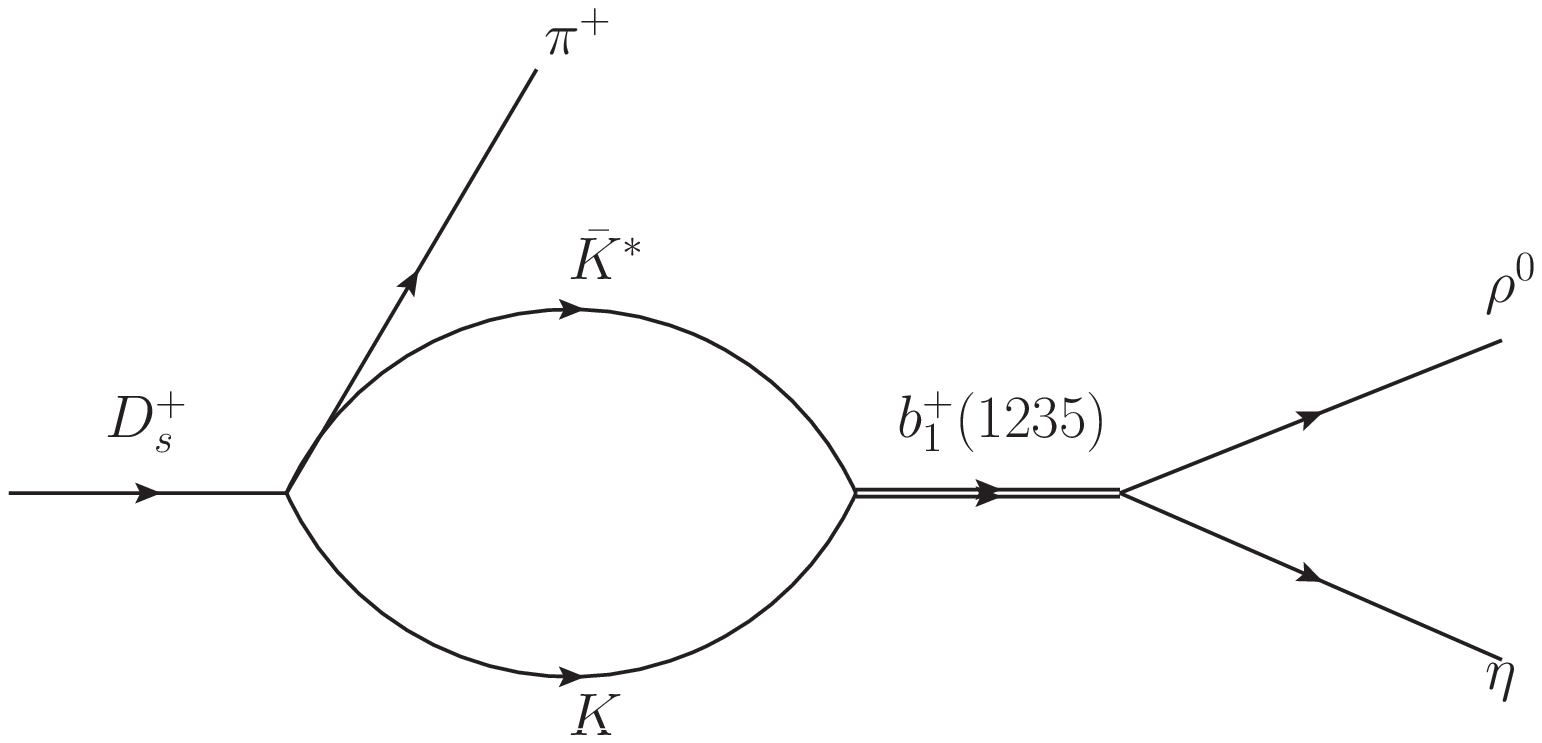}
  \includegraphics[width=0.4\textwidth]{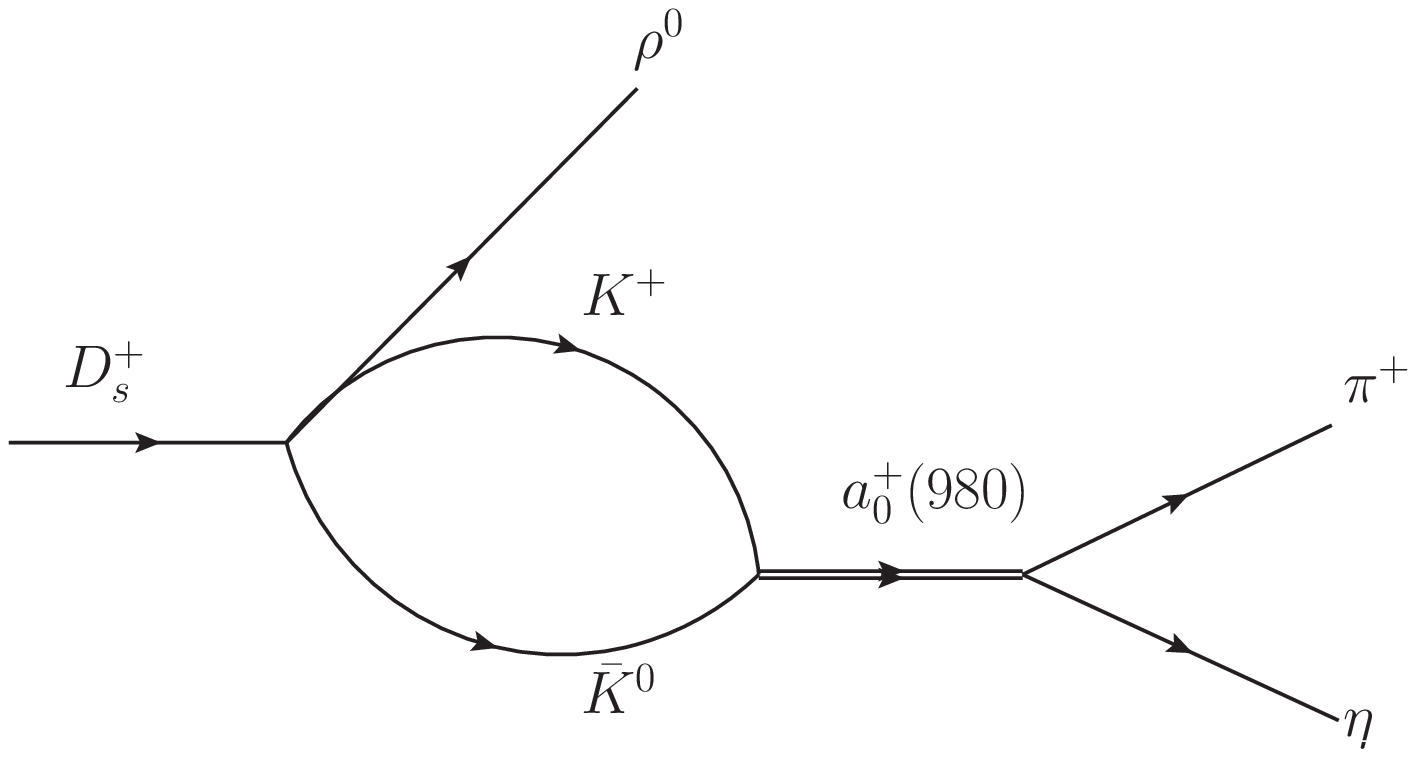}\qquad
  \caption{Diagrams stemming from the hadronization of one $q\bar{q}$ pair.}\label{fig3}
\end{figure}

In Ref.~\cite{Roca:2005nm,Oller:1997ti} the couplings of the resonances to the different components are given for the normalized states in isospin basis. We must obtain the projection of the states obtained here on the isospin states of~\cite{Roca:2005nm,Oller:1997ti}, which we do below.\\

a) $\bar{K}^{*0}K^{+}-K^{*+}\bar{K}^{0}$ has negative $G$-parity and $I=1$. It corresponds to the $a_1(1260)$ with $I_3=1$. Concretely,  using the convention of Ref.~\cite{Roca:2005nm}, we have 
\begin{align*}
    |a_1,I_3=1\rangle=\frac{1}{\sqrt{2}}(\bar{K}^{*0}K^{+}-K^{*+}\bar{K}^{0})
\end{align*}
Hence, 
\begin{align*}
   \langle a_1,I_3=1|\bar{K}^{*0}K^{+}-K^{*+}\bar{K}^{0}\rangle=\sqrt{2}. 
\end{align*}\\

b) ${K}^{*-}K^{+}-K^{*0}\bar{K}^{0}$ and $\bar{K}^{*0}K^{0}-K^{*+}{K}^{-}$ have both $G$-parity positive and belong to the $b_1$ resonance. Once again, with the convention of Ref.~\cite{Roca:2005nm} we have  
\begin{align*}
    |b_1,I_3=0\rangle=\frac{1}{\sqrt{2}}|\bar{K}^{*}K(I=1)+K^{*}\bar{K}(I=1)\rangle=\frac{1}{2}|\bar{K}^{*0}K^{0}-{K}^{*-}K^{+}-\bar{K}^{*+}K^{-}+K^{*0}\bar{K}^{0}\rangle
\end{align*}  
Hence,
\begin{align*}
    \langle b_1,I_3=0|{K}^{*-}K^{+}-K^{*0}\bar{K}^{0}\rangle=-1,~~~~
    \langle b_1,I_3=0|\bar{K}^{*0}K^{0}-{K}^{*+}K^{-}\rangle=1.
\end{align*}  \\

c) $K^{+}\bar{K}^{0}$ is the $I_3=1$ component of $K\bar{K}$ that couples to $a_0(980)$. Hence
\begin{align*}
    \langle a_0(980),I_3=1|K^{+}\bar{K}^{0}\rangle=1.
\end{align*}  \\
The resonances formed will decay to different channels, $a_1$  to $\rho^0\pi^{+}$, $b_1$ to $\rho^0\eta$ and $a_0$ to $\pi^{+}\eta$ and we have the picture depicted in Fig.~\ref{fig3}

We have established before the coupling of the $K^{*}\bar{K}$, $K\bar{K}$ channels to the $a_1$, $b_1$, and $a_0$ resonances. Now we must deal with their decays to $\rho\pi$, $\rho\eta$, and $\pi\eta$. We follow again Refs.~\cite{Roca:2005nm,Oller:1997ti}.\\

d) The $\rho\pi$ component for the $I_3=1$ state of the $a_1(1260)$ resonance is given considering the $\pi$, $\rho$ isospin multiples ($-\pi^{+}$, $\pi^{0}$, $\pi^{-}$), ($-\rho^{+}$, $\rho^{0}$, $\rho^{-}$) by
\begin{align*}
    |a_1,I_3=1\rangle=|-\frac{1}{\sqrt{2}}\rho^{+}\pi^{0}+\frac{1}{\sqrt{2}}\rho^{0}\pi^{+}\rangle
\end{align*}
and we are interested only in the second component, hence we have an overlap factor of $\frac{1}{\sqrt{2}}$.\\

e) The $I_3=0$ $\rho\eta$ component of the $b_1$ is 
\begin{align*}
    |b_1,I_3=0\rangle=|\rho^0\eta\rangle
\end{align*}
the overlap factor is $1$.\\

f) The $I_3=0$, $I_3=1$ components of the $a_0(980)$ are 
\begin{align*}
    | a_0,I_3=0\rangle=|\pi^0\eta\rangle,\,\,\,
    | a_0,I_3=1\rangle=-|\pi^+\eta\rangle
\end{align*}   
with all these weights calculated we obtain the following amplitude:
\begin{align}\label{18_1}
   t_{H1}(\pi^{+}\rho^0\eta)=  C\bigg[&-\sqrt{\frac{2}{3}}+ \frac{\eta}{\sqrt{3}}(1+\alpha+\beta)G_{K^{*}\bar{K}}(M_{\mathrm{inv}}(\rho^0\pi^{+}))\frac{g_{a_1, K^{*}\bar{K}}g_{a_1,\rho\pi}}{M^2_{\mathrm{inv}}(\rho^0\pi^{+})-M^2_{a_1}+iM_{a_1}\Gamma_{a_1}}\\\nonumber
        &- \sqrt{2}\beta G_{K\bar{K}}(M_{\mathrm{\mathrm{inv}}}(\pi^+\eta))\frac{g_{a_0, K\bar{K}}g_{a_0,\pi\eta}}{M^2_{\mathrm{\mathrm{inv}}}(\pi^+\eta)-M^2_{a_0}+iM_{a_0}\Gamma_{a_0}}\\\nonumber
        &- (\alpha+\gamma) G_{K^*\bar{K}}(M_{\mathrm{\mathrm{inv}}}(\rho^0\eta))\frac{g_{b_1, K^*\bar{K}}g_{b_1,\rho\eta}}{M^2_{\mathrm{\mathrm{inv}}}(\rho^0\eta)-M^2_{b_1}+iM_{b_1}\Gamma_{b_1}}
        \bigg]
\end{align}
where $G_{K^*\bar{K}}$, $G_{K\bar{K}}$ are the loop functions of two mesons which are regularized as in~\cite{Roca:2005nm} for $G_{K^*\bar{K}}$ and with a cut off method for $K\bar{K}$  as done in~\cite{Liang:2014tia,Xie:2014tma} with a cut off $600$~MeV. The couplings of the $a_1$, $b_1$ resonances are taken from~\cite{Roca:2005nm} and are shown in  Table~\ref{tableIII}, and the masses and widths from the PDG~\cite{ParticleDataGroup:2020ssz}. The $f_0(980)$ has a width of $10$-$100$~MeV in the PDG and we take $70$~MeV. One note is, however, mandatory here. The $a_0(980)$ usually is considered as a normal resonance. Yet, the high precision experiments where the $a_0(980)$ is seen lately~\cite{CLEO:2004umu,BESIII:2016tqo} show a shape of the $a_0$ as a strong sharp peak in the $\pi^0\eta$ mass distribution, typical of a cusp, corresponding to a barely failed state, or virtual state. This is also the case in a large number of theoretical works~\cite{Wu:2007jh,Hanhart:2007bd,Roca:2012cv,Xie:2014tma,Aceti:2012dj}. In this case the couplings are not well defined. In fact the couplings go to zero when one approaches a threshold as a consequence of the Weinberg compositeness condition~\cite{Weinberg:1965zz,Baru:2010ww,Kinugawa:2021ykv,Li:2021cue,Song:2022yvz,Albaladejo:2022sux,Bruns:2022hmb}. This is so for one channel, but it also holds for all couplings when using coupled channels when one approaches one threshold~\cite{Toki:2007ab,Gamermann:2009uq}. Because of this we replace in Eq.~(\ref{18_1})  
\begin{table}[H]
\centering
 \caption{The couplings of the $a_1$ and $b_1$ states in the unit of MeV~\cite{Roca:2005nm}.}\label{tableIII}
\setlength{\tabcolsep}{14pt}
\begin{tabular}{cccc}
\hline
\hline
\multicolumn{2}{c}{$a_1$} & \multicolumn{2}{c}{$b_1$}\\
\hline
$g_{\bar{K}^{*}K}$ & $g_{\rho\pi}$ & $g_{K^{*}\bar{K}}$ & $g_{\rho\eta}$\\
\hline
$1872-i1486$ & $-3.795+i2330$ & $-3041+i498$ & $6172-i75$\\
\hline
\hline
\end{tabular}
\end{table}

\begin{align}\label{19_1}
    \frac{g_{a_0, K\bar{K}}g_{a_0,\pi\eta}}{M^2_{\mathrm{inv}}(\pi^+\eta)-M^2_{a_0}+iM_{a_0}\Gamma_{a_0}} \to t^{I=1}_{K\bar{K},\pi\eta}
\end{align}
where $t^{I=1}_{K\bar{K},\pi\eta}$ is taken from the model of Ref.~\cite{Xie:2014tma} using the chiral unitary approach with the $\pi\eta$ and $K\bar{K}$ channels.

\subsection{Rescattering from the tree level $\eta\rho^0\pi^{+}$ component}
\begin{figure}[H]
  \centering
  \includegraphics[width=0.4\textwidth]{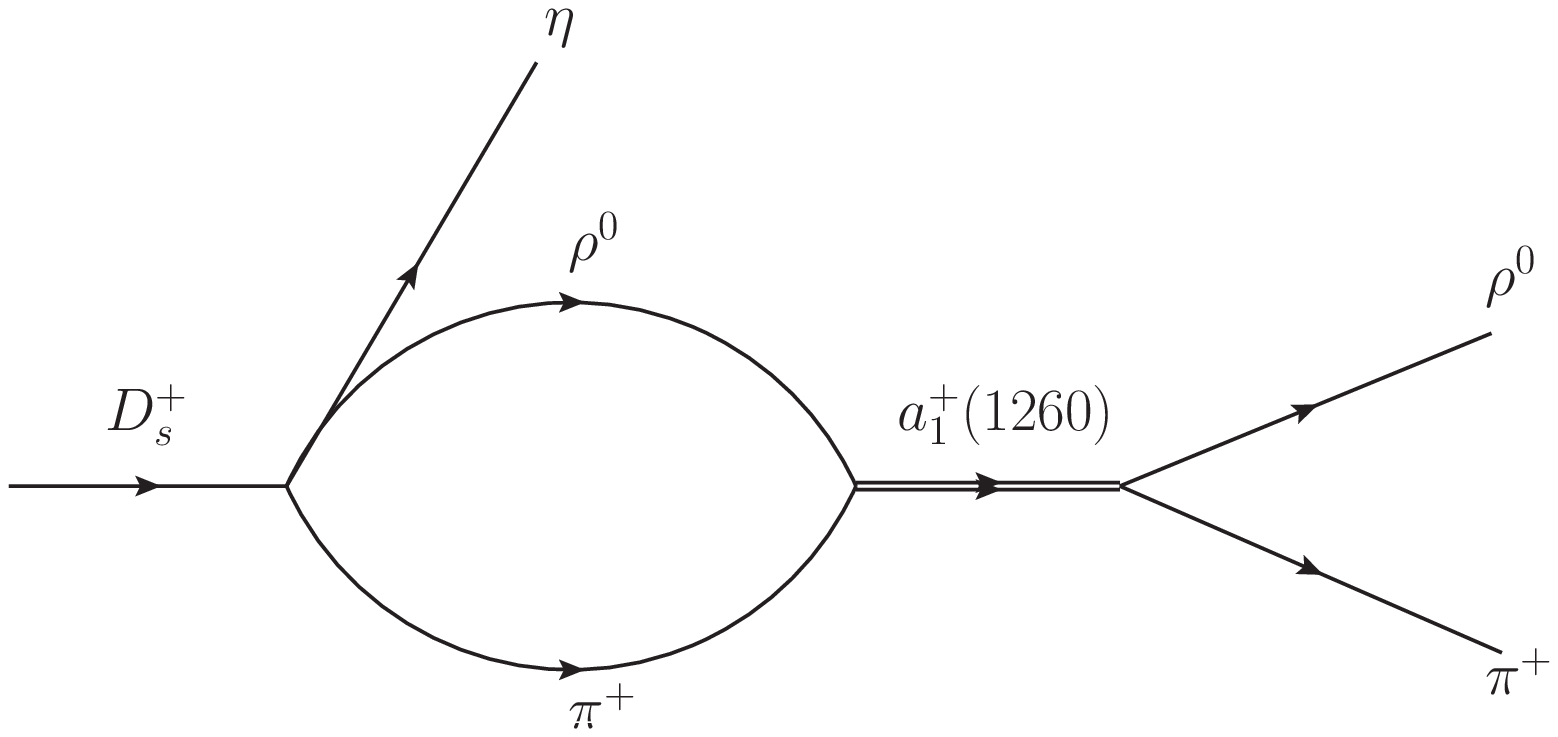}
  \includegraphics[width=0.4\textwidth]{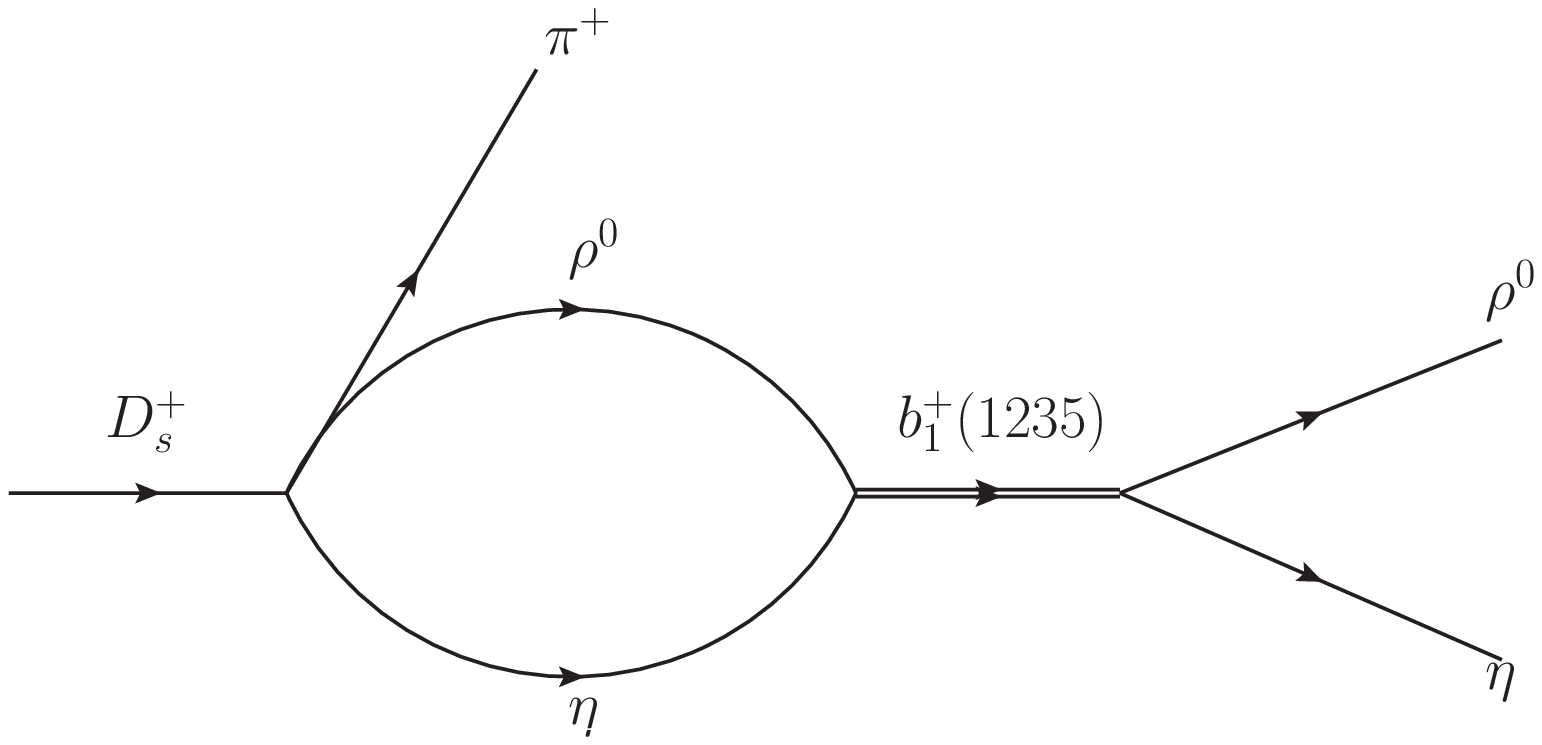}
  \includegraphics[width=0.4\textwidth]{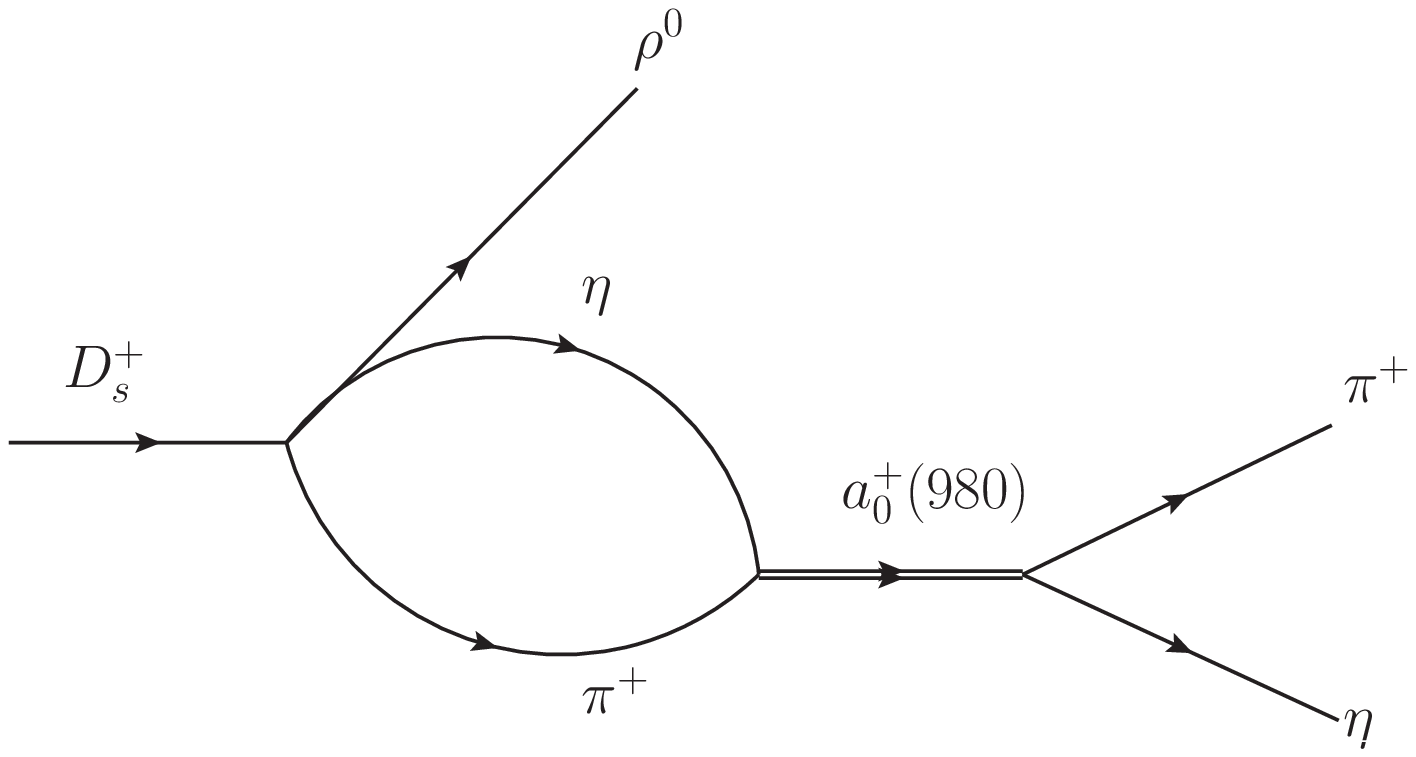}
  \caption{Diagrams from the rescattering of  pairs of mesons $\rho^0\pi^{+}\eta$ production.}\label{fig4}
\end{figure}

In Eq.~(\ref{18_1}) we have the tree level $\eta\rho^0\pi^{+}$ term, and all the others come from one transition from the primary meson states generated in the weak process and one hadronization. In line with this extra final state interaction of the $K^*\bar{K}$, $K\bar{K}$ components we address here the mechanisms of rescattering of the pairs of mesons in the tree level amplitude going to the same states. This leads to the diagrams shown in Fig.~\ref{fig4}.

With the ingredients before it is easy to write this amplitude as 
\begin{align}\label{32_33}
    t_{\mathrm{RES}}(\rho^0\pi^{+}\eta)=&-C\sqrt{\frac{2}{3}}G_{\rho\pi}(M_{\mathrm{inv}}(\rho^0\pi^+))\frac{\frac{1}{\sqrt{2}}g_{a_1,\rho\pi}\frac{1}{\sqrt{2}}g_{a_1,\rho\pi}}{M^2_{\mathrm{inv}}(\rho^0\pi^{+})-M^2_{a_1}+iM_{a_1}\Gamma_{a_1}}\\\nonumber
    &-C\sqrt{\frac{2}{3}}G_{\rho\eta}(M_{\mathrm{inv}}(\rho^0\eta))\frac{g_{b_1,\rho\eta}g_{b_1,\rho\eta}}{M^2_{\mathrm{inv}}(\rho^0\eta)-M^2_{b_1}+iM_{b_1}\Gamma_{b_1}}\\\nonumber
    &-C\sqrt{\frac{2}{3}}G_{\pi^{+}\eta}(M_{\mathrm{inv}}(\pi^{+}\eta))\frac{g_{a_0,\pi\eta}g_{a_0,\pi\eta}}{M^2_{\mathrm{inv}}(\pi\eta)-M^2_{a_0}+iM_{a_0}\Gamma_{a_0}}
\end{align}
 where, once again, in the last term we shall make the replacement of Eq.~(\ref{19_1}). 

There is still a bit extra work having to do with the $\rho$ production and its decay to $\pi^{+}\pi^{-}$. First we must contract the $\rho^0$ polarization vector with some momentum, since the $D_s$, $\eta$, $\pi$ have no spin. Given that $\pi$ and $\eta$ are produced on the same footing we could have $\epsilon_\mu(p_{\pi}+p_{\eta})^\mu$, or $\epsilon_\mu P_{D_S}^\mu$, but since $p_D=p_{\pi}+p_{\eta}+p_\rho$ and $\epsilon_\mu P_{D_S}^\mu=0$, these terms are equivalent and we take $\epsilon_\mu P_{D_S}^\mu$.

\begin{figure}[H]
  \centering
  \includegraphics[width=0.4\textwidth]{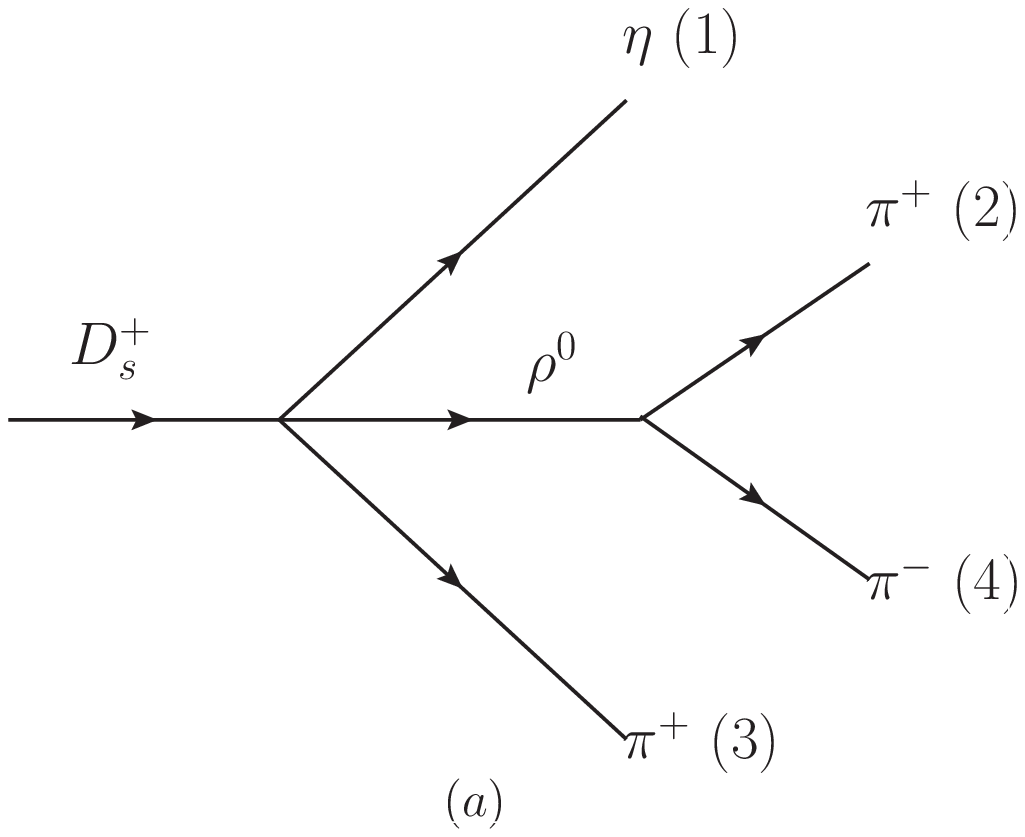}
  \includegraphics[width=0.4\textwidth]{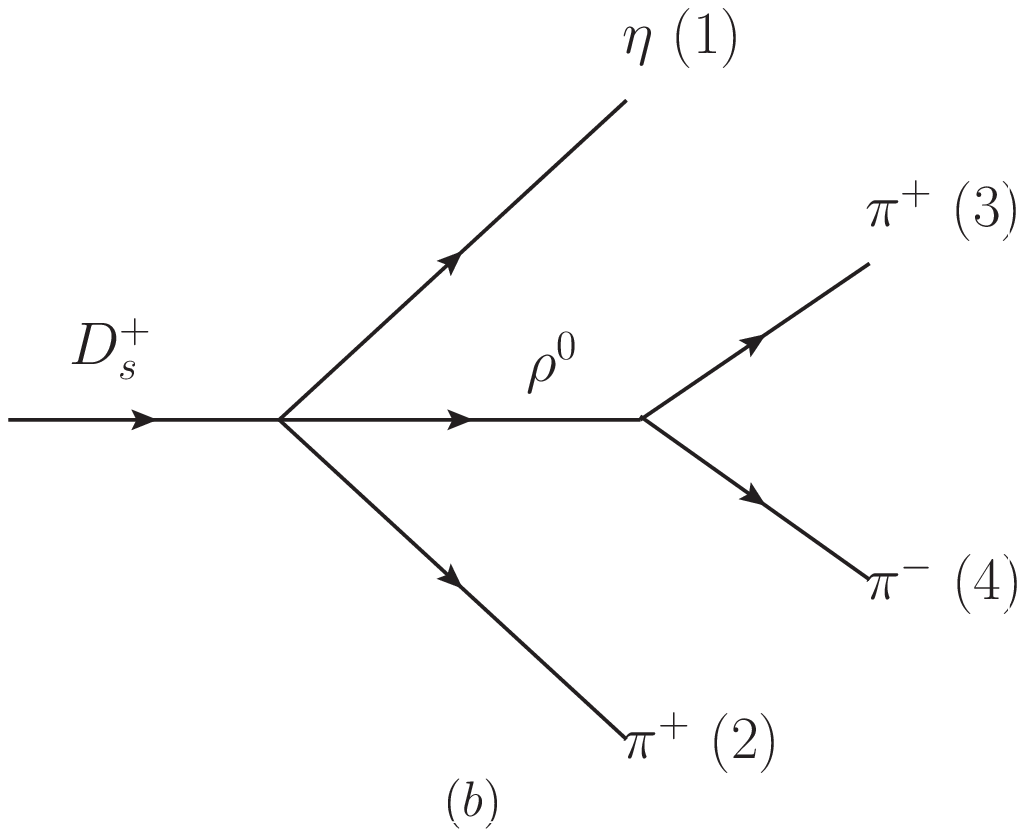}
  \caption{Symmetrized amplitude with $\rho$ decaying to two pions.}\label{fig5}
\end{figure}

Next, when the $\rho^0$ decays to $\pi^{+}\pi^{-}$ there are two $\pi^{+}$ at the end, and one must symmetrize the amplitude. We shall have two diagrams, as depicted in fig.~\ref{fig5}.

Then, taking for instance the diagram of Fig.~\ref{fig5}(a), we would have for the $\rho$ propagator the sum over the $\rho$ polarization.  
\begin{align*}
    \sum_{pol}P_D^\mu\epsilon_\mu\epsilon_\nu(p_4-p_2)^{\nu}=P_D^\mu(-g_{\mu\nu}+\frac{q_\mu q_\nu}{m_{\rho^2}})(p_4-p_2)^\nu
\end{align*}
with $q=p_2+p_4$. Since $q(p_2-p_4)=m_{\pi}^2-m_{\pi}^2=0$ we have then $-p_D^\mu(p_4-p_2)_\mu$ and the amplitudes become 
\begin{align}\label{21_1}
    t_{\rho}=-C\Bigg[
    &{P}_{D_{s}}\cdot({p}_{4}-{p}_{2})\frac{1}{{M^2_{\mathrm{inv}}(\rho,a)}-M^2_{\rho}+iM_{\rho}\Gamma_{\rho}}t^{(a)}\\\nonumber
    &+{P}_{D_{s}}\cdot({p}_{4}-{p}_{3})\frac{1}{{M^2_{\mathrm{inv}}(\rho,b)}-M^2_{\rho}+iM_{\rho}\Gamma_{\rho}}t^{(b)}
    \Bigg]
\end{align}
where $t^{(a)}$ and $t^{(b)}$ are the amplitudes evaluated before with the momenta configuration of the diagrams of Fig.~\ref{fig5} (a) and (b). $M^2_{\mathrm{inv}}(\rho,a)$ and $M^2_{\mathrm{inv}}(\rho,b)$ are $p_\rho^2$ for the configurations of Fig.~\ref{fig5} (a) and (b) respectively. We have omitted the $\rho\pi\pi$  coupling which is incorporated in the $C$ global coefficient.

\subsection{Two hadronizations with external emission: $f_0(980)$ contribution}
\begin{figure}[H]
  \centering
  \includegraphics[width=0.45\textwidth]{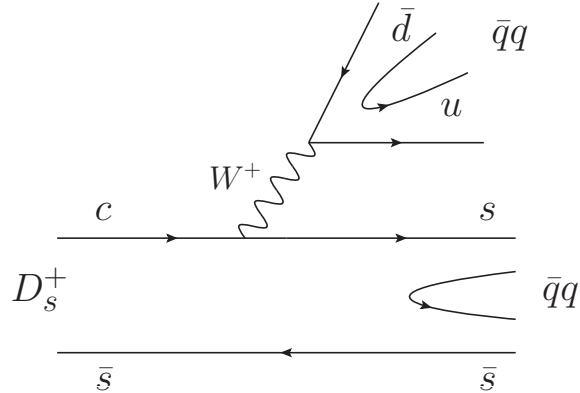}
  \caption{Mechanism with two hadronization producing four pseudoscalars.}\label{fig6}
\end{figure}

\begin{figure}
  \centering
  \includegraphics[width=0.4\textwidth]{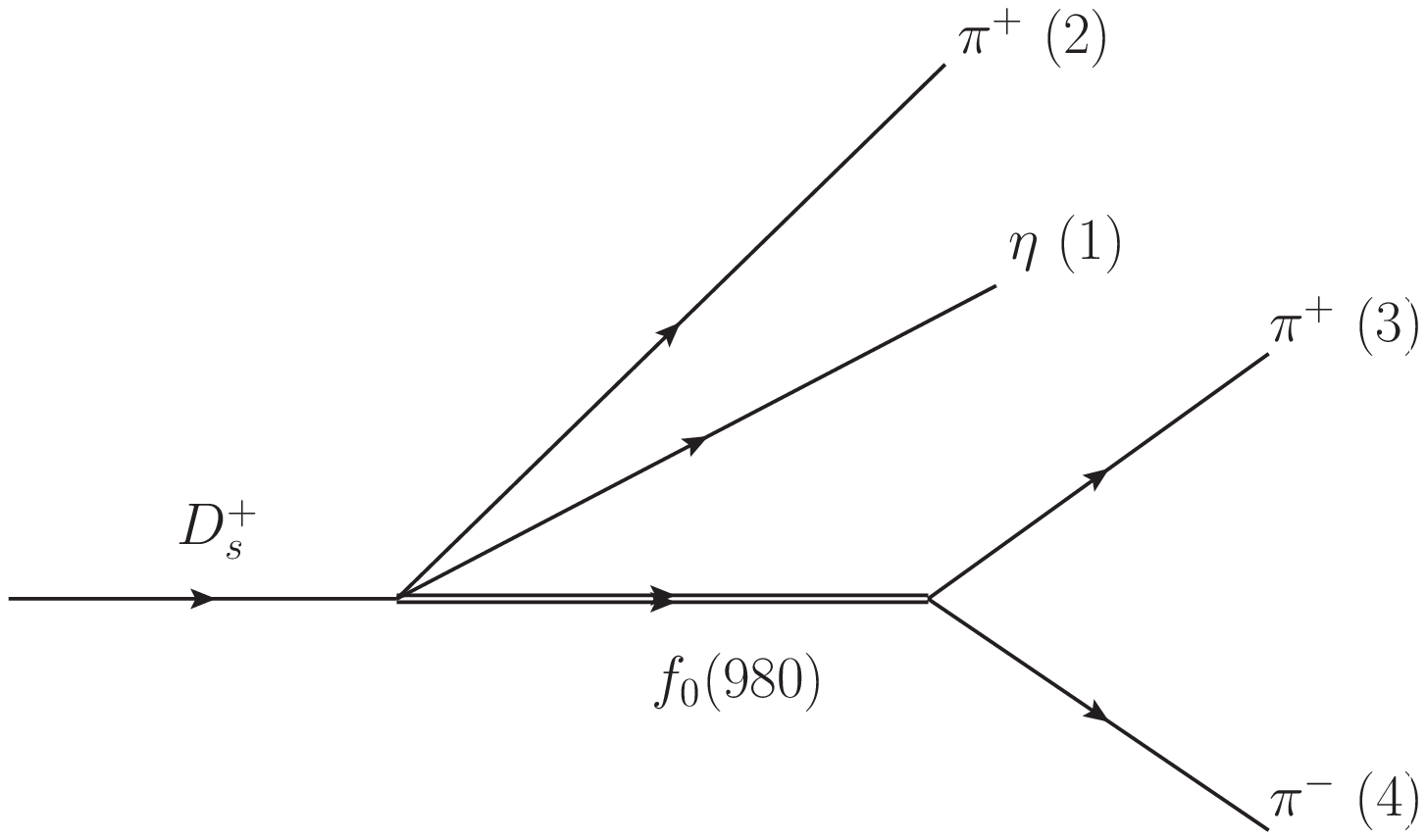}
  \includegraphics[width=0.4\textwidth]{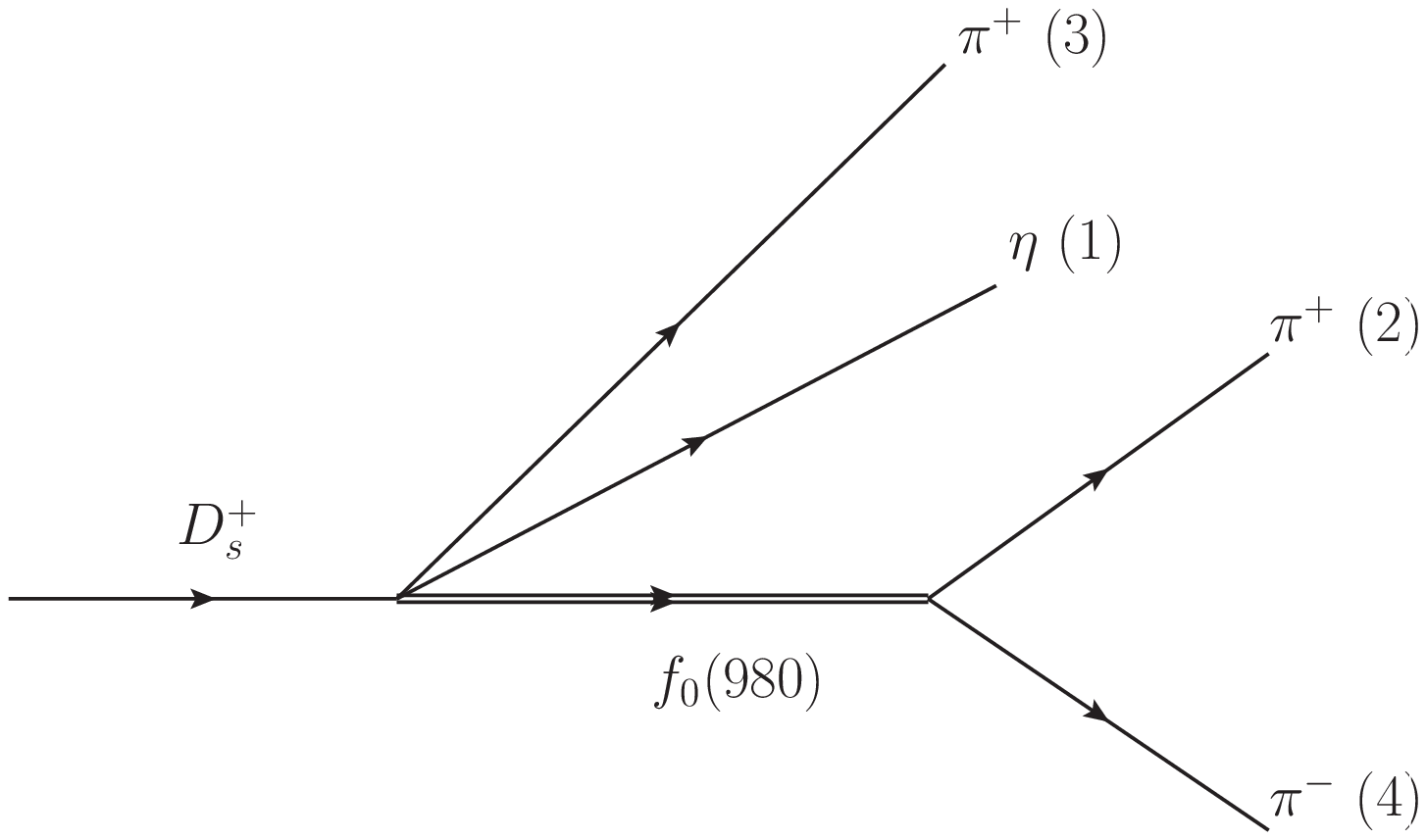}
  \caption{Diagrams stemming from two hadronizations producing the $f_0(980)$ resonance.}\label{fig7}
\end{figure}

In the $\pi^{+}\pi^{-}$ mass distribution of Ref.~\cite{BESIII:2021aza} one observes two structures, one for the $\rho^0$ and another one for the $f_0(980)$. So far neither the $f_0(980)$ nor the $a_0(980)$, with $I_3=-1$ ($\pi^-\eta$), have  appeared in our scheme. The reason lies in the fact that, up to now, we have only considered mechanism with one hadronization providing $VPP$. Now we consider two hadronizations leading to four pseudoscalars $PPPP$. The corresponding external emission mechanism proceeds through the diagram given in Fig.~\ref{fig6}
Since $\eta=\frac{1}{\sqrt{3}}(u\bar{u}+d\bar{d})-\frac{1}{\sqrt{3}}s\bar{s}$ with the mixing of~\cite{Bramon:1992kr}, the hadronization of the $\bar{d}u$ component can lead to $\pi^+\eta$ while the hadronization of $s\bar{s}$ can lead $K\bar{K}$. We thus obtain $\pi^+\eta K\bar{K}$ which is not the desired final state. However we can have the $K\bar{K}\to\pi^+\pi^-$ transition and then we get the $\pi^+\pi^+\pi^-\eta$ final state. In terms of resonances, $K\bar{K}$ coming from the $I=0$ $s\bar{s}$ pairs, can only give rise to the $f_0(980)$, the coupling to the $f_0(500)$ being very small. Hence, we find a mechanism for the production of this resonance, and we have two new diagrams, which are depicted in Fig.~\ref{fig7}.

A caveat  must be  recalled at this point, since as discussed in~\cite{Sun:2015uva} the coupling of $W^+$ to two pseudoscalars goes as the difference of energies of the two pseudoscalars which will vanish in the $W^+$ rest frame for two particles with the same mass. However, $\eta$ and $\pi$ have very different masses and the term survives.

\begin{figure}[H]
  \centering
    \includegraphics[width=0.4\textwidth]{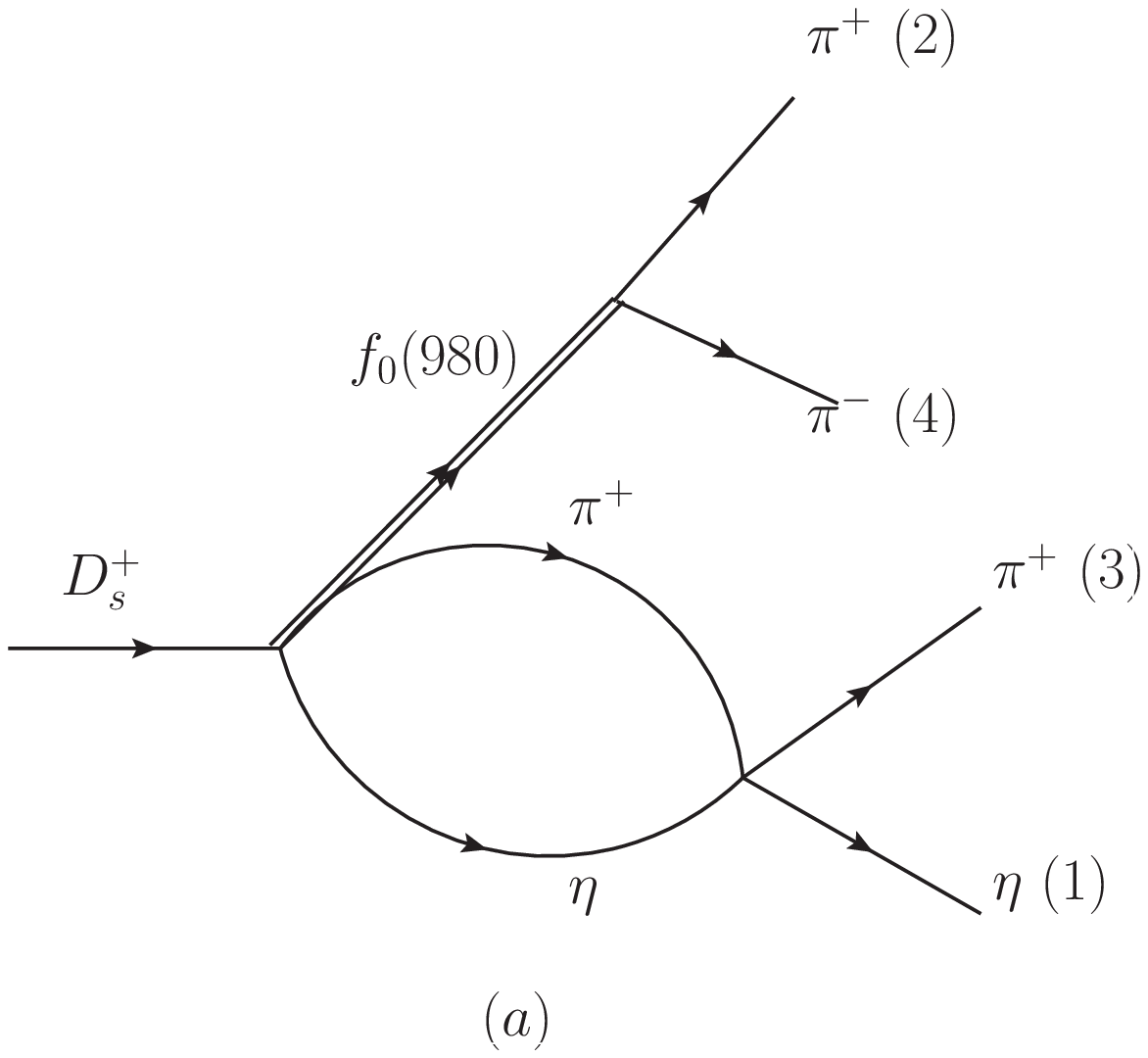}
    \includegraphics[width=0.4\textwidth]{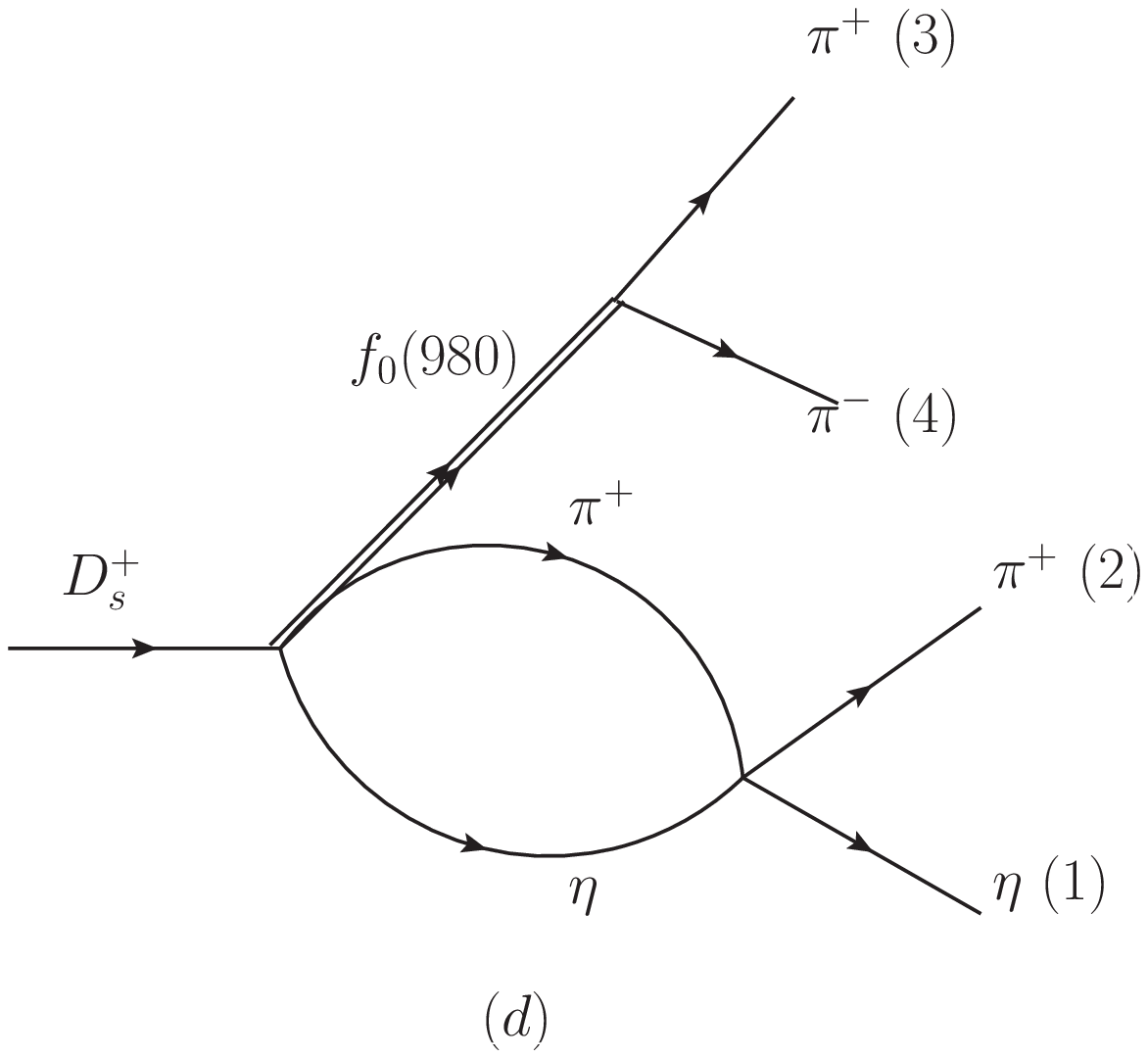}  
    \includegraphics[width=0.4\textwidth]{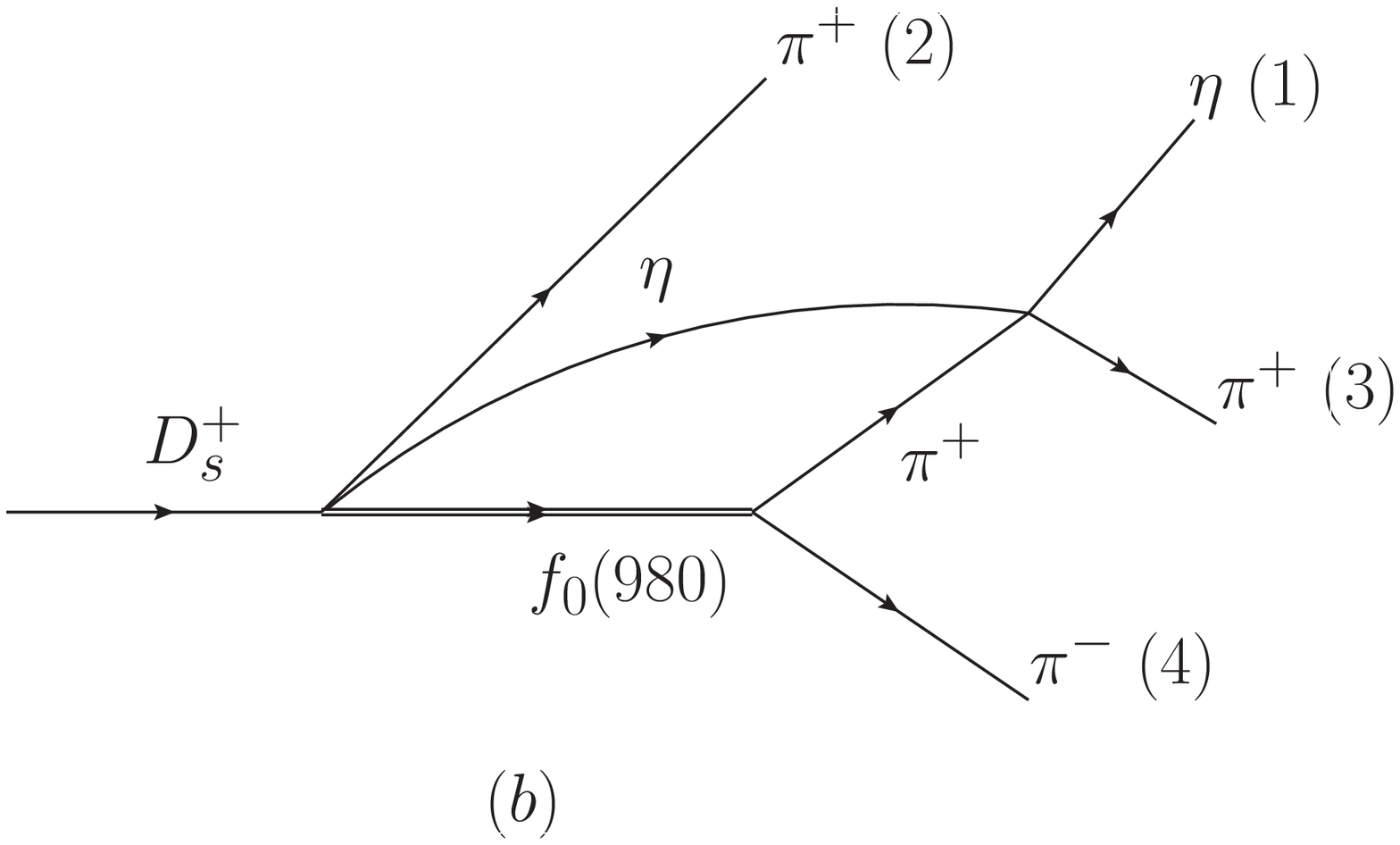}
    \includegraphics[width=0.4\textwidth]{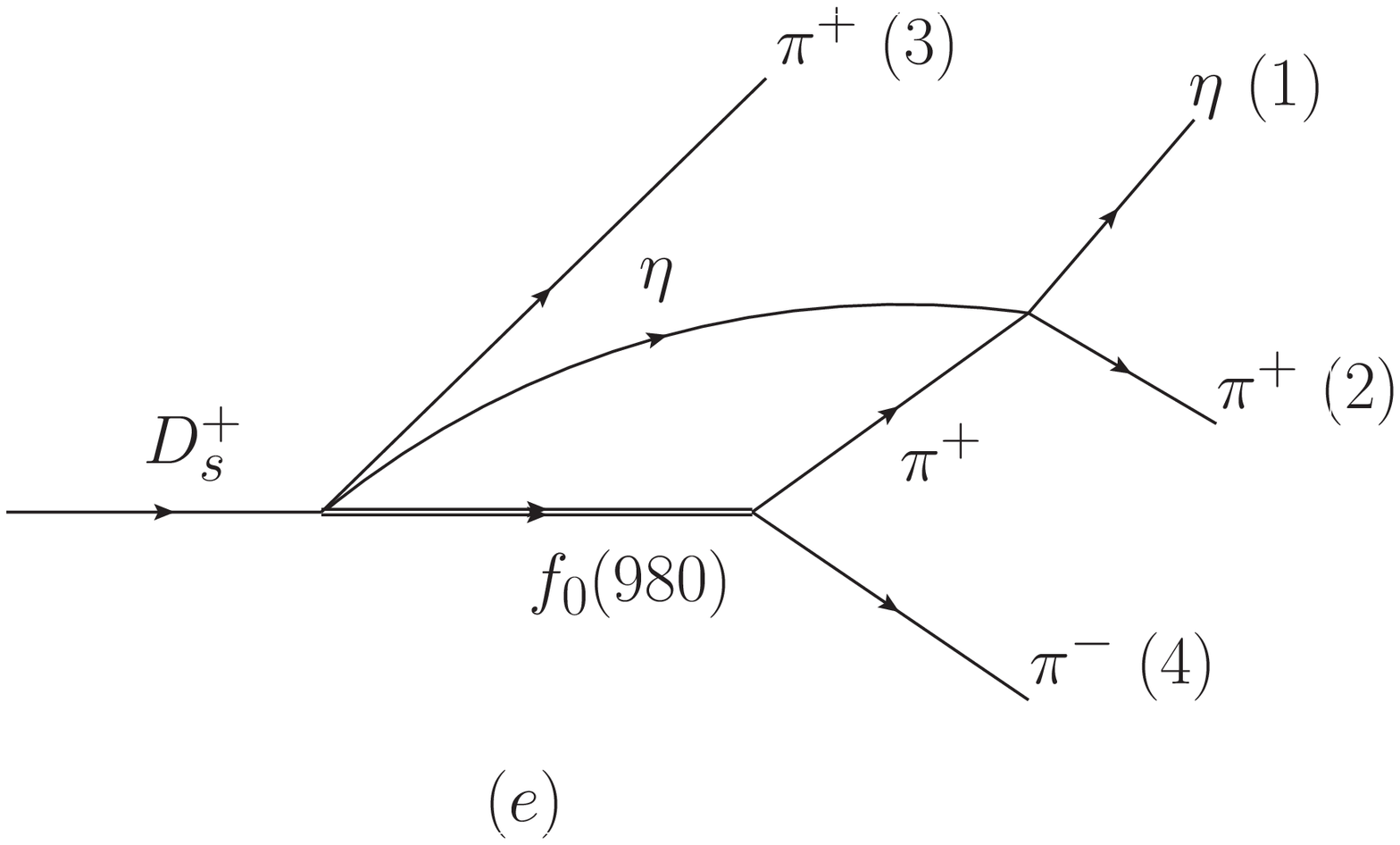}
    \includegraphics[width=0.4\textwidth]{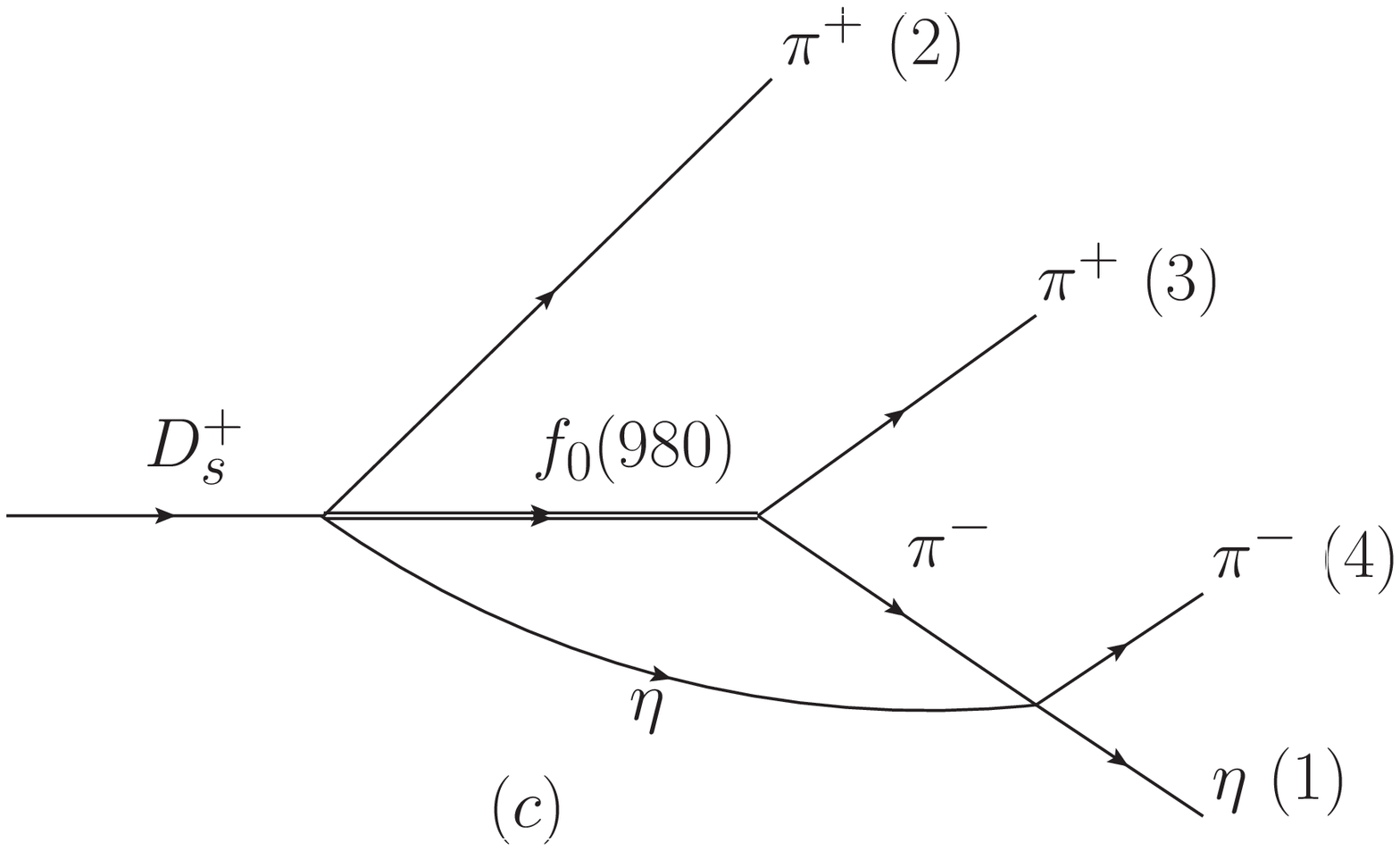}
    \includegraphics[width=0.4\textwidth]{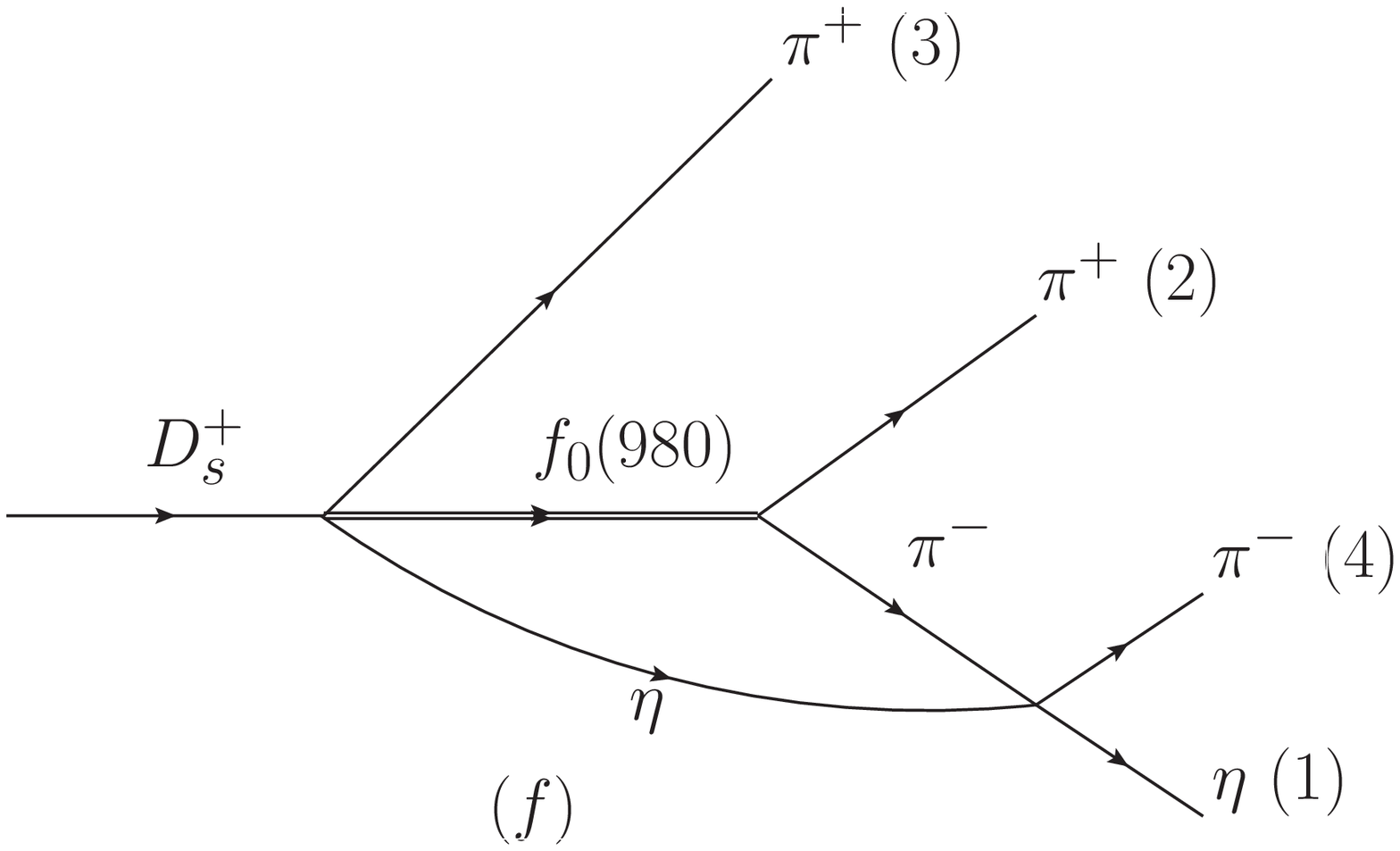}
  \caption{Diagrams stemming from those of Fig.~\ref{fig7} after rescattering of two psedoscalars.}\label{fig8}
\end{figure}

After generating the two new mechanisms of Fig.~\ref{fig7} we can proceed further and take into account the final state interaction of the different components. This is depicted in Fig.~\ref{fig8}. We should note that all the couplings here are of $S$-wave nature unlike the former diagrams that all contained a $\rho^0$ decaying to $\pi^+\pi^-$ in $P$-wave. A similar diagram to Fig.~\ref{fig8} (b) could be done replacing the $f_0(980)$ by the $\rho$ meson, but the extra loop with respect to the former diagrams, together with the factor $(p_4-p_2)^0$ from Eq.~(\ref{21_1}) does not make it competitive in comparison with the mechanisms of Fig.~\ref{fig8} and we do not consider it here.

In the evaluation of the diagrams in Fig.~\ref{fig8} where the $f_0(980)$ is in the loop we make some approximation to evaluate it. What makes the approximation good is the realization that the $f_0(980)$ in the loop is very far off shell. One interesting way to see it is to test how far one is from having a Triangle Singularity~\cite{Landau:1959fi} which would place the $\pi\eta$ and $f_0(980)$ simultaneously on mass shell. For this we apply Eq.~(18) of Ref.~\cite{Bayar:2016ftu} and see that one is far from satisfying that condition, and since the $\pi\eta$ can be obviously on shell, from where the loop gets its maximum contribution, the $f_0(980)$ will be off shell. This allows us to factorize the $f_0(980)$ propagator in the loop. Taking as an example the loop in Fig.~\ref{fig8} (f) we would have 
\begin{align*}
    D_{f_0}=&\frac{1}{M^2_{\mathrm{inv}(\pi^{-}_\mathrm{int}\pi^{+}\sj{(2)})}-M^2_{f_0}+iM_{f_0}\Gamma_{f_0}}
\end{align*}  
with $\pi^{-}_\mathrm{int}$, the $\pi^-$ inside the loop, where
\begin{align}\label{24_1}
    M^2_{\mathrm{inv}}(\pi^{-}_\mathrm{int}\pi^{+}\sj{(2)})=&(w_{\pi^{+}\sj{(2)}}+w_{\pi^{-}_\mathrm{int}})^2-(\textbf{p}_{\pi^{+}\sj{(2)}}+\textbf{p}_{\pi^{-}_\mathrm{int}})^2\\\nonumber
       =&2m^2_{\pi^{+}}+2w_{\pi^{+}\sj{(2)}}w_{\pi^{-}_\mathrm{int}}-2\textbf{p}_{\pi^{+}\sj{(2)}}\textbf{p}_{\pi^{-}_\mathrm{int}}\\\nonumber
       \approx &2m^2_{\pi}+2w_{\pi^{+}\sj{(2)}}w_{\pi^{-}_\mathrm{int}}
\end{align}
where in the last equation we have removed the $\textbf{p}_{\pi^{+}\sj{(2)}}\textbf{p}_{\pi^{-}_\mathrm{int}}$ term, a sensible approximation if we evaluate Eq.~(\ref{24_1}) in the $\pi^-\eta$ rest frame when performing the $\textbf{p}_{\pi^{-}_\mathrm{int}}$ integration in the loop. Recalling that we get most of the contribution to the loop when $\pi^{-}_\mathrm{int}\eta$ are on shell we get 
\begin{align}
    w_{\pi^{-}_\mathrm{int}}=&\frac{M^2_{\mathrm{inv}}(\pi^{-}\eta)+m^2_{\pi^{-}}-m^2_{\eta}}{2M_{\mathrm{inv}}(\pi^{-}\eta)},\\\nonumber
    w_{\pi^{+}\sj{(2)}}=&\frac{p_{\pi^{+}\sj{(2)}}\cdot(p_{\pi^{-}}+p_{\eta})}{M_{\mathrm{inv}}(\pi^{-}\eta)}.
\end{align}
We multiply $D_{f_0}$ by a factor $M_{f_0}\Gamma_{f_0}$ to have a dimensionless magnitude 
\begin{align*}
    \Tilde{D}_{f_0}=M_{f_0}\Gamma_{f_0}D_{f_0}
\end{align*}
and then we get for the diagrams of Fig.~\ref{fig7} and Fig.~\ref{fig8}, $t_1(f_0)$ and $t_2(f_0)$ given by 
\begin{align}\label{25_1}
    t_1(f_0)&=C\mu\bigg[
    \Tilde{D}_{f_0}(M_{\mathrm{inv}}\bigg(\pi^{+}(3)\pi^{-})\bigg)+\Tilde{D}_{f_0}\bigg(M_{\mathrm{inv}}(\pi^{+}(2)\pi^{-})\bigg)
\bigg]
\end{align} 
\begin{align}\label{25_2}
    t_2(f_0)&=t_\mathrm{2a}(f_0)+t_\mathrm{2b}(f_0)+t_\mathrm{2c}(f_0)+t_\mathrm{2d}(f_0)+t_\mathrm{2e}(f_0)+t_\mathrm{2f}(f_0)
\end{align}  
with 
\begin{align}\label{25_3}
    t_\mathrm{2a}(f_0)=& C\mu \Tilde{D}_{f_0}\bigg(M_{\mathrm{inv}}(\pi^{+}\sj{(2)}\pi^{-})\bigg)G_{\pi\eta}\bigg(M_{\mathrm{inv}}(\pi^{+}\sj{(3)}\eta)\bigg)t_{\pi^{+}\eta,\pi^{+}\eta}\bigg(M_{\mathrm{inv}}(\pi^{+}\sj{(3)}\eta)\bigg)\\\nonumber
    t_\mathrm{2d}(f_0)=& C\mu \Tilde{D}_{f_0}\bigg(M_{\mathrm{inv}}(\pi^{+}\sj{(3)}\pi^{-})\bigg)G_{\pi\eta}\bigg(M_{\mathrm{inv}}(\pi^{+}\sj{(2)}\eta)\bigg)t_{\pi^{+}\eta,\pi^{+}\eta}\bigg(M_{\mathrm{inv}}(\pi^{+}\sj{(2)}\eta)\bigg)\\\nonumber
    t_\mathrm{2b}(f_0)=& C\mu \Tilde{D}_{f_0}\bigg(M_{\mathrm{inv}}(\pi^{-}\pi^{+}_{\mathrm{int}})\bigg)G_{\pi\eta}\bigg(M_{\mathrm{inv}}(\pi^{+}\sj{(3)}\eta)\bigg)t_{\pi^{+}\eta,\pi^{+}\eta}\bigg(M_{\mathrm{inv}}(\pi^{+}\sj{(3)}\eta)\bigg)\\\nonumber
    t_\mathrm{2c}(f_0)=& C\mu \Tilde{D}_{f_0}\bigg(M_{\mathrm{inv}}(\pi^{+}\sj{(3)}\pi^{-}_\mathrm{int})\bigg)G_{\pi\eta}\bigg(M_{\mathrm{inv}}(\pi^{-}\eta)\bigg)t_{\pi^{-}\eta,\pi^{-}\eta}\bigg(M_{\mathrm{inv}}(\pi^{-}\eta)\bigg)\\\nonumber
    t_\mathrm{2e}(f_0)=&  C\mu\Tilde{D}_{f_0}\bigg(M_{\mathrm{inv}}(\pi^{-}\pi^{+}_\mathrm{int})\bigg)G_{\pi\eta}\bigg(M_{\mathrm{inv}}(\pi^{+}\sj{(2)}\eta)\bigg)t_{\pi^{+}\eta,\pi^{+}\eta}\bigg(M_{\mathrm{inv}}(\pi^{+}\sj{(2)}\eta)\bigg)\\\nonumber
    t_\mathrm{2f}(f_0)=& C\mu \Tilde{D}_{f_0}\bigg(M_{\mathrm{inv}}(\pi^{+}\sj{(2)}\pi^{-}_\mathrm{int})\bigg)G_{\pi\eta}\bigg(M_{\mathrm{inv}}(\pi^{-}\eta)\bigg)t_{\pi^{-}\eta,\pi^{-}\eta}\bigg(M_{\mathrm{inv}}(\pi^{-}\eta)\bigg)
\end{align} 

and summing all them we have 
\begin{align}\label{25_4}
    t_{f_0}=t_{1}(f_0)+t_{2}(f_0)
\end{align}
We should note that through the mechanism of Fig.~\ref{fig8} ($c$) ($f$) we obtain the for first time a signal for the $a_0^-\to\pi^-\eta$, which is also clearly visible in the $\pi^-\eta$ invariant mass of Ref.~\cite{BESIII:2021aza}.
\subsection{Two hadronizatons with internal emission}
\begin{figure}[H]
  \centering
   \includegraphics[width=0.45\textwidth]{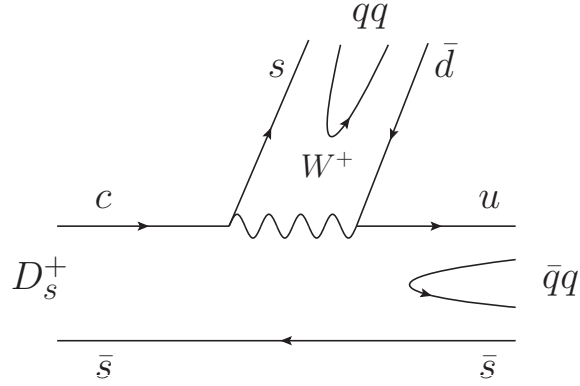}
  \caption{Diagram for internal emission with two hadronizations.}\label{fig9}
\end{figure}

\begin{figure}[H]
  \centering
  \includegraphics[width=0.45\textwidth]{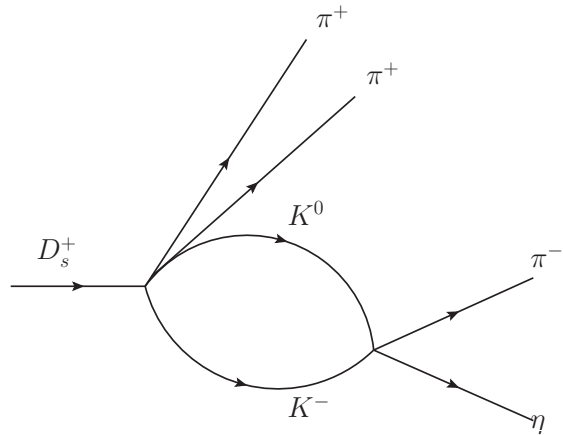}
  \caption{Mechanism contributing to the  $\pi^{+}\pi^{+}\pi^{-}\eta$ production stemming from the diagram of Fig.~\ref{fig9}.}\label{fig10}
\end{figure}

Now we produce directly four pseudoscalars through the mechanism of Fig.~(\ref{fig9})
The hadronization produces now
\begin{align}
    s\bar{d} \to &(P^2)_{32}=K^{-}\pi^{+}-\bar{K}^0\frac{\pi^{0}}{\sqrt{2}}\\\nonumber
    u\bar{s} \to &(P^2)_{13}=\frac{\pi^{0}}{\sqrt{2}}K^{+}+\pi^{+}K^0\\\nonumber
\end{align}
and thus, we have together

\begin{align*}
K^{-}K^{+}\pi^{+}\frac{\pi^{0}}{\sqrt{2}}+K^{-}\pi^{+}\pi^{+}K^0-\bar{K}^0\frac{\pi^{0}}{\sqrt{2}}\frac{\pi^{0}}{\sqrt{2}}K^{+}-\bar{K}^0\frac{\pi^{0}}{\sqrt{2}}\pi^{+}K^0
\end{align*}
which indicates that only the second term can contribute to our process through $K^-K^0\to a_0^-\to\pi^-\eta$ via the loop shown in Fig.~\ref{fig10}
The amplitude for this process will be written as 
\begin{align}
    t_\mathrm{DIE}=C\nu G_{K\bar{K}}\bigg(M_{\mathrm{\mathrm{inv}}}(\pi^{-}\eta)\bigg)t_{\pi^-\eta,K^0K^-}
    \end{align}
where $t_{\pi^-\eta,K^0K^-}\equiv t_{\pi\eta,\pi\eta}^{I=1}$.

Coming from a double hadronization and internal emission, the term should be smaller than former ones and we refrain from studying further interactions. We collect all the amplitudes from $t_{H1}$, $t_{RES}$, properly symmetrized and accounting for $\rho$ decay as shown in Eq.~(\ref{21_1}) plus $t_{f_0}$ and $t_\mathrm{DIE}$ to construct the full amplitude, $t$. We have $5$ parameters in addition to the global normalization constant $C$, which will be fitted to the data. In the following section we show how to evaluate the different cross sections.

\section{evaluation of the differential cross section}
The width for the $D_s$ decay into $\pi^{+}\pi^{+}\pi^{-}\eta$ is given by
\begin{align}
    \Gamma=\frac{1}{2M_{D_s}}\frac{1}{2}\int\frac{\mathrm{d}^{3}p_{\eta}}{(2\pi)^3}\frac{1}{2E_{\eta}}\int\frac{\mathrm{d}^{3}p_{\pi^{+}}}{(2\pi)^3}\frac{1}{2E_{\pi^{+}}}\int\frac{\mathrm{d}^{3}p'_{\pi^{+}}}{(2\pi)^3}\frac{1}{2E'_{\pi^{+}}}\int\frac{\mathrm{d}^{3}p_{\pi^{-}}}{(2\pi)^3}\frac{1}{2E_{\pi^{-}}}(2\pi)^4\delta^{(4)}(P-p_{\eta}-p_{\pi^{+}}-p'_{\pi^{+}}-p_{\pi^{-}})|t|^{2},
\end{align} 
with the factor $1/2$ since we have a symmetrized amplitude with respect to the two $\pi^+$. We kill the $p_{\pi^{-}}$ integration with the $\delta^3()$ function, which gives us 
\begin{align*}
    \textbf{p}_{\pi^{-}}=-(\textbf{p}_{\eta}+\textbf{p}_{\pi^{+}}+\textbf{p}'_{\pi^{+}}).
\end{align*}
We introduce the variable 
\begin{align*}
    \textbf{P}_{\pi}=&\textbf{p}_{\pi^{+}}+\textbf{p}'_{\pi^{+}},\nonumber\\
    \textbf{q}_{\pi}=&\textbf{p}_{\pi^{+}}-\textbf{p}'_{\pi^{+}},
\end{align*} 
which gives us  
\begin{align}
    \Gamma=\frac{1}{2M_{D_s}}\frac{1}{2}\frac{1}{2}\int\frac{\mathrm{d}^{3}p_{\eta}}{(2\pi)^3}\frac{1}{2E_{\eta}}\int\frac{\mathrm{d}^{3}P_{\pi}}{(2\pi)^3}\int\frac{\mathrm{d}^{3}q}{(2\pi)^3}\frac{1}{2E_{\pi^{+}}}\frac{1}{2E'_{\pi^{+}}}\frac{1}{2E_{\pi^{-}}}(2\pi)\delta\bigg(M_{D_s} - E_\eta - E_{\pi^{+}} - E'_{\pi^{+}}-\sqrt{m_{\pi^-}^2+(\textbf{P}_\pi+\textbf{p}_\eta)^2}\bigg)|t|^{2}.
\end{align} 
The $\delta()$ condition allows us to obtain $\cos{\theta}$ between $\textbf{P}_{\pi}$ and $\textbf{p}_{\eta}$ as a function of the other variables and we choose $\eta$ in the $z$ direction. We find 
\begin{align*}
    \cos{\theta}\equiv A=\frac{1}{2{P}_\pi p_\eta}\bigg[(M_{D_s} - E_\eta - E_{\pi^{+}} - E'_{\pi^{+}})^2-m_{\pi^-}^2-\textbf{P}_\pi^2-\textbf{p}_\eta^2\bigg]
\end{align*} 
but we must demand that $|\cos{\theta}|\leq1$ and we implement the factor in the integrand
\begin{align*}
    \theta(1-A^2)\theta(M_{D_s} - E_\eta - E_{\pi^{+}} - E'_{\pi^{+}}).
\end{align*} 
Then the width is finally written as 
\begin{align}
    \Gamma=\frac{1}{8M_{D_s}}\frac{1}{\pi}\int{p^2_{\eta}\mathrm{d}p_{\eta}}\frac{1}{2E_{\eta}}\int\frac{P^2_{\pi}\mathrm{d}P_{\pi}\mathrm{d}{\phi}}{(2\pi)^3}\int\frac{\mathrm{d}^3q}{(2\pi)^3}\frac{1}{2E_{\pi^{+}}}\frac{1}{2E'_{\pi^{+}}}\frac{1}{2P_{\pi}p_{\eta}}|t|^{2}\theta(1-A^2)\theta(M_{D_s} - E_\eta - E_{\pi^{+}} - E'_{\pi^{+}}),       
\end{align}
and we take
\begin{align}\label{rotation_10}
\textbf{p}_{\eta}=p_{\eta} 
    \left(
  \begin{array}{ccc}
0\\
0\\
1\\
  \end{array}
\right);~~
\textbf{P}_{\pi}=P_{\pi} 
    \left(
  \begin{array}{ccc}
\sin{\theta}\cos{\phi}\\
\sin{\theta}\sin{\phi}\\
\cos{\theta}  \\
  \end{array}
\right),~~
(\cos{\theta}= A,~~\sin{\theta}=\sqrt{1-A^2}).
\end{align}

However, it is convenient to define $\textbf{q}$ with respect to $\textbf{P}_{\pi}$ for integration purposes since it allows $A\equiv\cos{\theta}$  to be expressed in terms of the integration variables, as
$\textbf{p}_{\pi^+}=\frac{1}{2}(\textbf{P}_{\pi}+\textbf{q})$, $\textbf{p}'_{\pi^+}=\frac{1}{2}(\textbf{P}_{\pi}-\textbf{q})$. Hence we define $\Tilde{\textbf{q}}$ related to $\textbf{P}_\pi$ as the $z$ axis as 
\begin{align*}
\Tilde{\textbf{q}}=q 
    \left(
  \begin{array}{ccc}
\sin{\Tilde{\theta}_{q}}\cos{\Tilde{\phi}_{q}}\\
\sin{\Tilde{\theta}_{q}}\sin{\Tilde{\phi}_{q}}\\
\cos{\Tilde{\theta}_{q}}  \\
  \end{array}
\right),
\end{align*} 
and to write it in the $D_s$ rest frame with $\textbf{p}_\eta$ in the $z$ direction, we make two rotations and find $\textbf{q}=R\Tilde{\textbf{q}}$ with
\begin{align*}
R=R_{\phi}R_{\theta}=
    \left(
  \begin{array}{ccc}
\cos{\phi} & -\sin{\phi} & 0\\
\sin{\phi} & \cos{\phi} & 0\\
0 & 0 & 1 \\
  \end{array}
\right)
 \left(
  \begin{array}{ccc}
\cos{\theta} & 0 & \sin{\theta} \\
0 & 1 & 0\\
-\sin{\theta} & 0 & \cos{\theta} \\
  \end{array}
\right)=
 \left(
  \begin{array}{ccc}
\cos{\phi}\cos{\theta} & -\sin{\phi} & \cos{\phi}\sin{\theta} \\
\sin{\phi}\cos{\theta} & \cos{\phi} & \sin{\phi}\sin{\theta}\\
-\sin{\theta} & 0 & \cos{\theta} \\
  \end{array}
\right)
\end{align*}
With this choice: $E_{\pi^{+}}+E'_{\pi^{+}}$ entering the definition of  $A\equiv\cos{\theta}$ is given by
\begin{align*}
    E_{\pi^{+}}+E'_{\pi^{+}}  &= \sqrt{m_\pi^2+\frac{1}{4}(\textbf{P}_\pi+\textbf{q})^2}+\sqrt{m_\pi^2+\frac{1}{4}(\textbf{P}_\pi-\textbf{q})^2}\\\nonumber
    &=\sqrt{m_\pi^2+\frac{1}{4}P^2_\pi+\frac{1}{4}q^2+\frac{1}{2}P_\pi q\cos\Tilde{\theta}_{q}}+\sqrt{m_\pi^2+\frac{1}{4}P^2_\pi+\frac{1}{4}q^2-\frac{1}{2}P_\pi q\cos\Tilde{\theta}_{q}}
\end{align*}
With these new variables and using $p_\eta \mathrm{d}p_\eta=E_\eta \mathrm{d} E_\eta$ we can write the width as :
\begin{align}\label{30_1}
    \Gamma=\frac{1}{8M_{D_s}}\frac{1}{4\pi}\int~\mathrm{d}E_{\eta}\int\frac{P_{\pi}\mathrm{d}P_{\pi}\mathrm{d}{\phi}}{(2\pi)^3}\int\frac{q^2\mathrm{d}q\mathrm{d}\cos{\Tilde{\theta}_{q}}\mathrm{d}{\Tilde{\phi}_{q}}}{(2\pi)^3}\frac{1}{2E_{\pi^{+}}}\frac{1}{2E'_{\pi^{+}}}|t|^{2}\theta(1-A^2)\theta(M_{D_s} - E_\eta - E_{\pi^{+}} - E'_{\pi^{+}}).         
\end{align}
The choice of variable was done such that we can evaluate the integral and mass distributions using Monte Carlo integration. We have $6$ integration variables. Random values within limits are chosen for all of them and the events are weighed by the integrand, and the $\theta()$ functions in Eq.~(\ref{30_1}) determine the phase space. With these variables, we construct all the momenta of the $4$ final particles, which allows us to calculate the six invariant mass distributions. We accumulate the weighed events in boxes of mass distributions of $25$~MeV, like in the experiment, and with $\sim 10^7$ generated events we get very accurate numerical mass distributions.

\section{results}
\begin{figure}[H]
  \centering
  \includegraphics[width=0.9\textwidth]{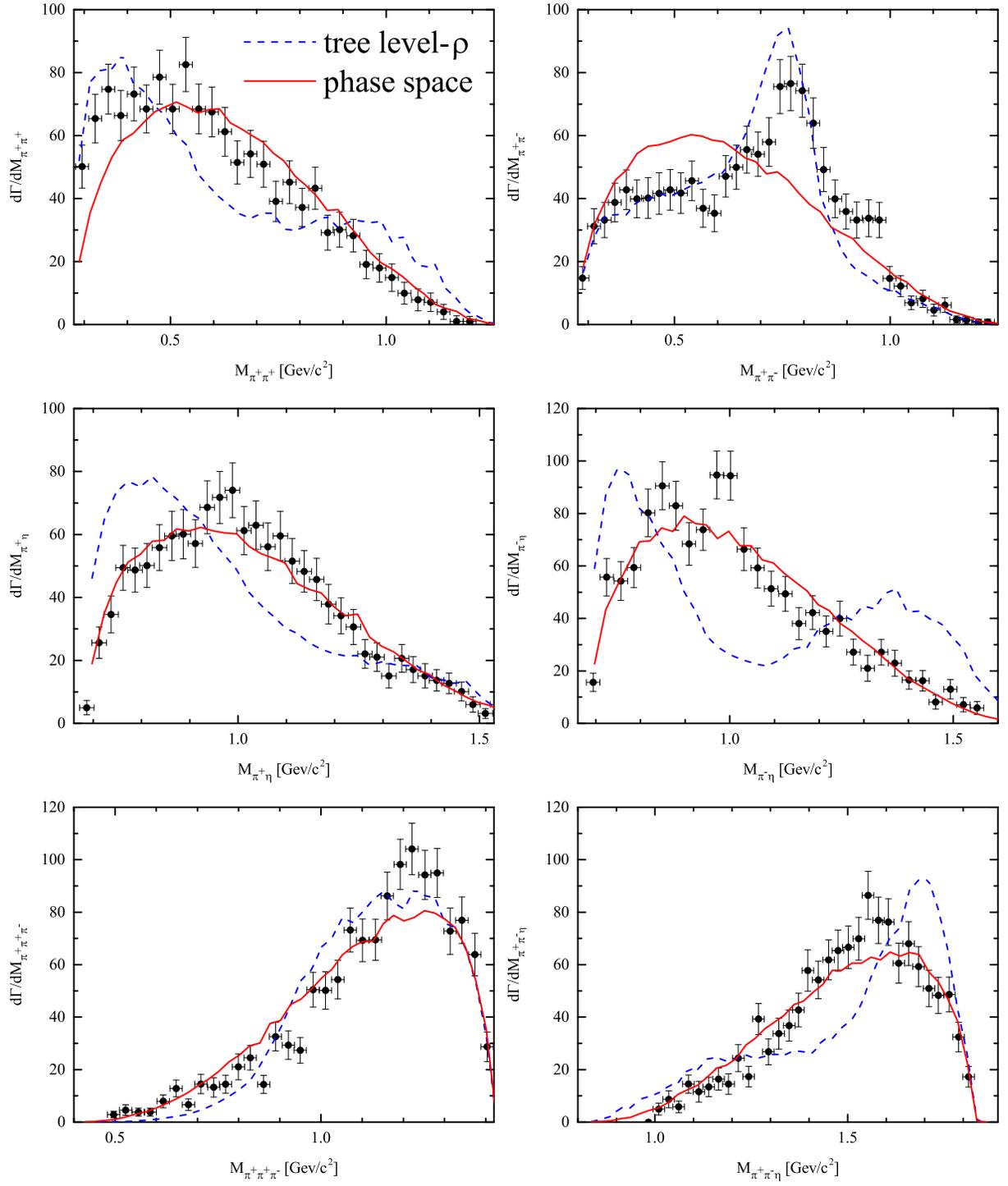}
  \caption{Phase space and the tree-level of $\rho$.}\label{fig11}
\end{figure}

First we go to Eq.~(\ref{30_1}) and substitute $t=1$ in order to obtain the mass distributions with pure phase space. The results are shown, with the mass distributions, normalized to the data, in Fig.~\ref{fig11}. In the same figure we also show the results obtained at the tree level of $t_{H1}$, with the $\pi^+\rho\eta$ amplitude of Eq.~(\ref{18_1}) keeping only the term $\sqrt{\frac{2}{3}}$, considering the $\rho$ decay as shown in Eq.~(\ref{21_1}).

What we see in the figure is that in the case of phase space, one is obviously missing all the different structures which are visible in the experimental data: the $\rho^0$ peak in the $M(\pi^+\pi^-)$ distribution, the $a_0^+(980)$ in the $M(\pi^+\eta)$ distribution, and the $f_0(980)$ in the $M(\pi^+\pi^-)$ distribution. The effect of the $a_1(1260)$ (very wide) and of the $b_1$are not so clear. Yet, the phase space alone fairly reproduces the gross features of the mass distribution, expect for the case of the $M(\pi^+\pi^-)$, where the prominent peak of the $\rho$ is obviously absent.

In the same figure we have the contribution of the $\pi^+\rho^0\eta$ term alone at the tree level. We can now see that the $M(\pi^+\pi^-)$ distribution is fairly well reproduced with its prominent $\rho$ peak. It is very interesting to note that this term alone also creates a broad bump in $M(\pi^+\pi^-)$ at low energies around $0.5$~GeV. This is particularly relevant since one could intuitively think that this bump comes from the production of the $f_0(500)$, as is seen in many other experiments. But we saw that in our exhaustive list of mechanisms of the reaction the $f_0(500)$ was never produced. We obtained the $f_0(980)$ via the $K\bar{K}$ coupling, but the $f_0(500)$ is well known to couple extremely weakly to this component~\cite{Oller:1997ti,Kaiser:1998fi,Locher:1997gr,Nieves:1999bx}. Instead, we see that without producing this resonance, the $\rho$ term gives rise to this wide bump. This is a consequence of the fact that we have two $\pi^+$ and if one  $\pi^+\pi^-$ pair creates the $\rho^0$, the other $\pi^+\pi^-$ pair creates a different structure, which in this case is a $M(\pi^+\pi^-)$ distribution mimicking the $f_0(500)$ distribution. The other interesting thing is that now this term has widely distorted the other mass distributions, creating peaks or bumps that are in sheer contradiction with the experiment. These extra peaks are also well known in mass distributions when one has many particles in the final state, and are called replicas or reflections in other channels of genuine resonances of one channel. The fact that we also have two $\pi^+$ certainly has something to say. What is clear after we introduce the important $\rho^0\pi^+\eta$ term of tree level, is that we need other contributions to describe the experimental data.

Our formalism comes from the systematic consideration of all possible mechanisms and we saw that they indeed produced the $a_0^+$, $a_0^-$, $a_1^+$, $b_1^+$, $f_0(980)$ resonances through rescattering of meson-meson components that were produced in a first step of the weak reaction upon the hadronization of one or two pairs of $q\bar{q}$. These transitions had no freedom since for them we took the amplitudes generated by the chiral unitary approach. At the end of the large amount of terms collected we had, up to a global normalization constant, $5$ free parameters, We should emphasize that the filter of $G$-parity negative eliminated terms, and it was linear contributions of states produced in different mechanisms that at the end gave us the desired $\pi^+\pi^+\pi^-\eta$ in the final state. Also, some terms of $G$-parity negative were shown not to lead to the desired final state and were eliminated. At the end, it is not easy to trace back the relative size of these parameters which have been assumed to be real as it would be the case in a quark model evaluation of the weak process plus the subsequent hadronization. Note, however, that the $G$ functions and $t_i$ matrices are complex, therefore our final amplitudes are complex and there are inevitably large interferences.

\begin{figure}[H]
  \centering
  \includegraphics[width=0.9\textwidth]{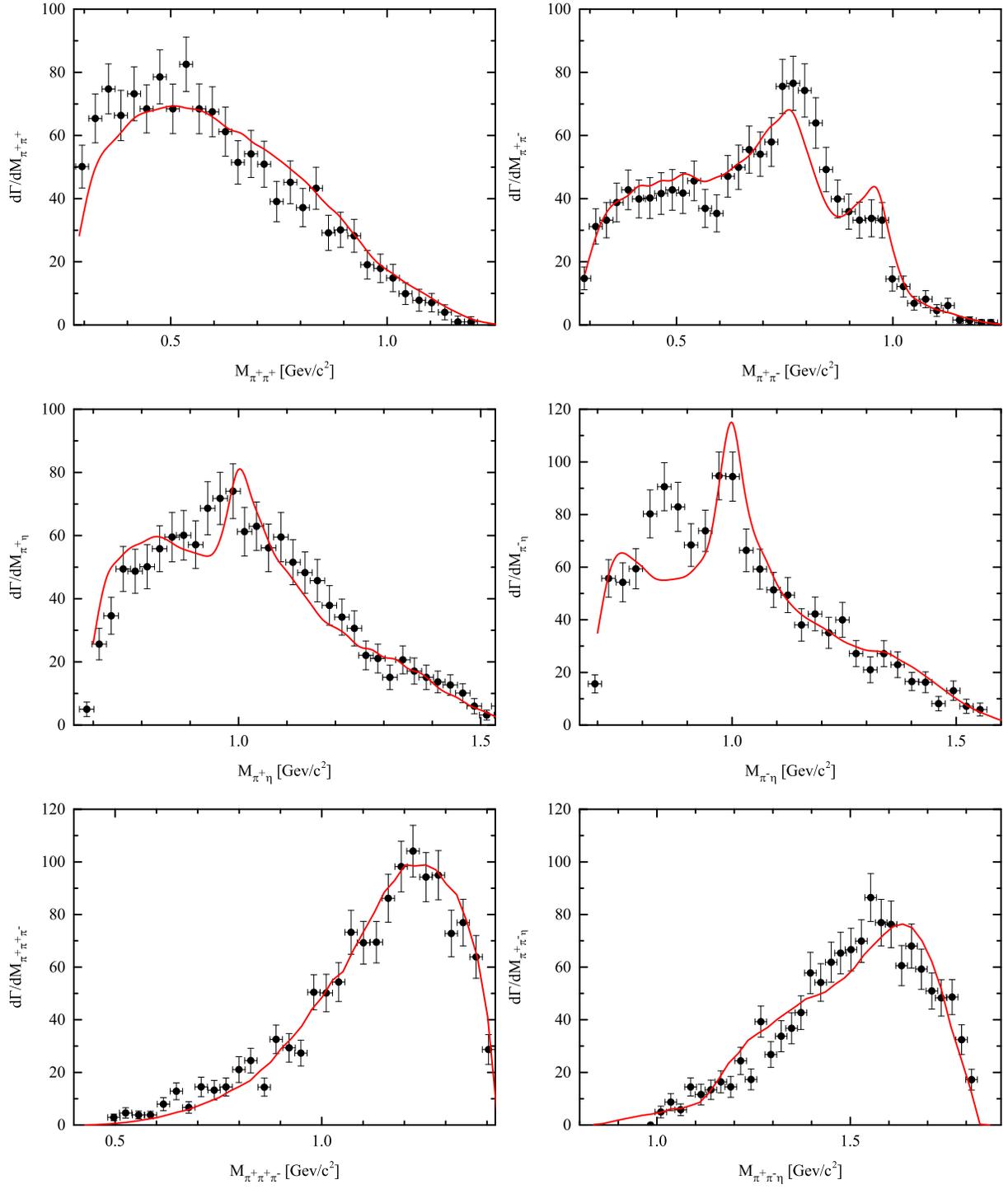}
  \caption{The experimental data from BESIII~\cite{BESIII:2021aza}. The red lines present the different mass distribute of $\pi^{+}\pi^{+}$, $\pi^{+}\pi^{-}$, $\pi^{+}\eta$, $\pi^{-}\eta$,  $\pi^{+}\pi^{+}\pi^{-}$, and $\pi^{+}\pi^{-}\eta$, individually. }\label{fig12}
\end{figure}

We, thus, conduct a best fit to all the six mass distributions and obtain the results that we show in Fig.~\ref{fig12}. The improvement over the mass distributions of the tree level $\rho^0\pi^+\eta$ term is remarkable. The $\chi_{d.o.f.}^2$ is $1.77$ that can be considered acceptable. The values obtained for the parameters are shown below
\begin{align*}
    \alpha=4.2,~~\beta=2.9,~~\gamma=-3.9,~~\mu=-31.2,~~\nu=39.1.
\end{align*}
Inspection of the figures shows a rather featureless $M(\pi^+\pi^+)$  distributions as also seen in the experiment. The $M(\pi^+\pi^-)$  distribution shows a clearly the $\rho^0$ and the $f_0(980)$ peaks. Also the low energy bump of the mass distribution is well reproduced which, as we discussed above, should not be associated to $f_0(500)$ excitation. In the same $M(\pi^+\pi^-)$  distribution we also see now a peak for the $f_0(980)$. The $M(\pi^+\eta)$ distribution comes out fairly well and a peak is seen corresponding to the $a_0^+(980)$ resonance. The resulting $M(\pi^+\pi^+\pi^-)$ and  $M(\pi^+\pi^-\eta)$ distributions are also in good agreement with the corresponding experimental ones. The  $M(\pi^-\eta)$ distribution shows a clear peak for the $a^-_0(980)$, as in the experiment, and the low and high energy parts of the spectrum are also well reproduced. There is a discrepancy with the data in a peak around  $0.85$~GeV that we cannot reproduce and do not know its dynamical origin.

\begin{figure}[H]
  \centering
  \includegraphics[width=0.9\textwidth]{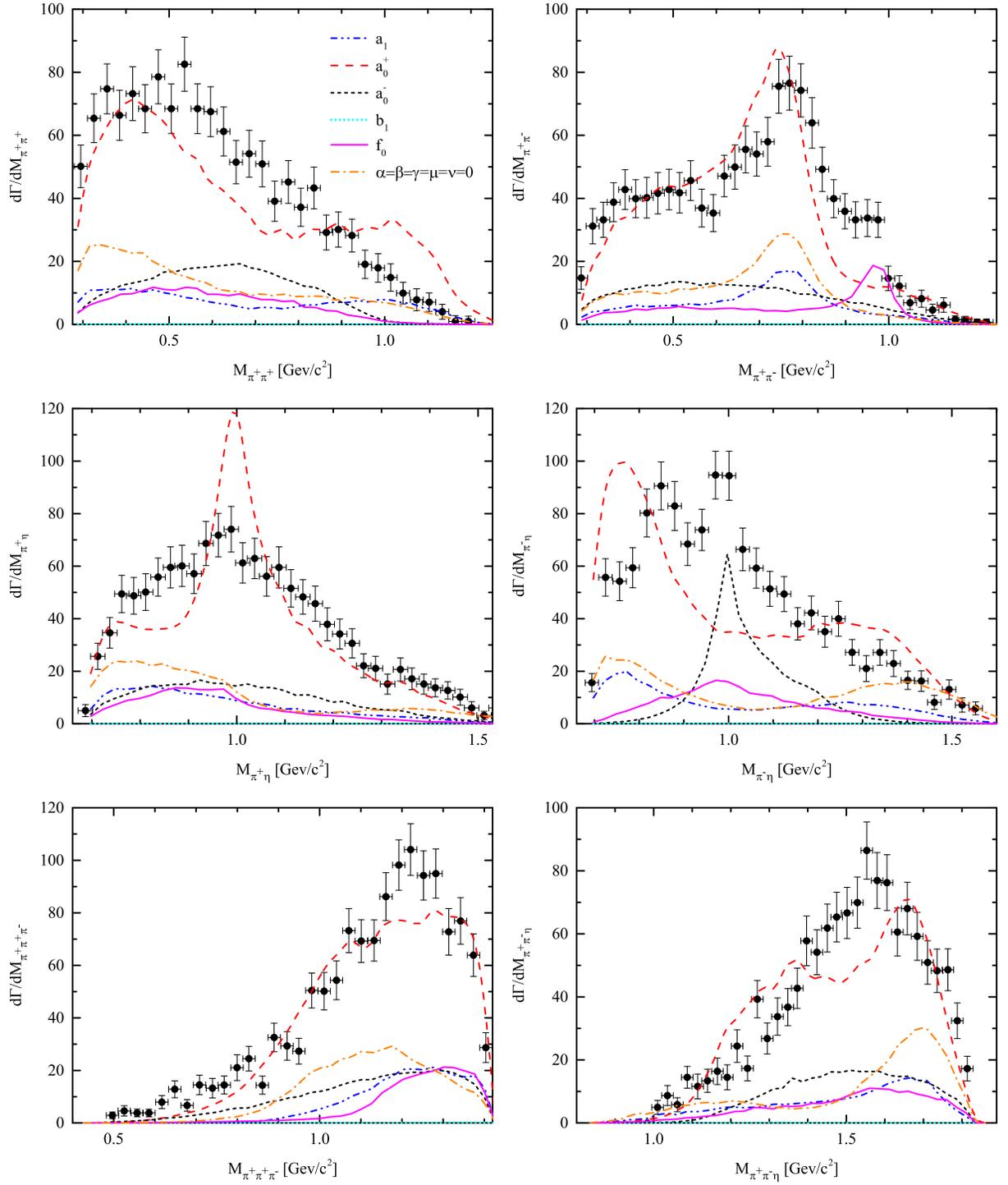}
  \caption{ The contributions of $a_1$, $a^+_0$, $a^-_0$, $b_1$, $f_0$, and external emission, respectively.}\label{fig13}
\end{figure}

Next, apart from the $\rho$ already addressed, we would like to discuss the contribution of the different resonances to the six invariant mass distributions in Fig.~\ref{fig13}. One must look at this information with care because there are interferences among the different amplitudes, but it gives an idea on how they appear in the process. For this we take the fitted amplitude and isolate the different terms of this amplitude where each resonance appear, and with only these terms we see the individual contributions to the mass distributions. As can be immediately appreciated, the dominant contribution comes from $a_0^+$. However, one should note that, as seen in Eq.~(\ref{18_1}) where the $a_0^+$ appears, it goes together with the $\rho^0$. Then it is not surprising to see a structure in  $M(\pi^+\pi^-)$ and $M(\pi^-\eta)$ similar to the one produced by the $\rho$ term alone shown in Fig.~\ref{fig11}. In our formalism we cannot disentangle that structure. The $a_0^-$ excitation has also a relatively large strength. The $a_1^+$ shows a clear peak around $1260$~MeV, in the $\pi^+\pi^+\pi^-$, as it should be. Curiously, our mechanisms allow the excitation of the $b_1^+$, but the strength obtained is practically negligible. One should note that in the experimental analysis the $b_1\pi$ mode was also not reported. Finally, we also show the contribution of the case where we set $\alpha=\beta=\gamma=\mu=\nu=0$. The remaining amplitude still contains the tree level $\pi^+\rho^0\eta$  of Eq.~(\ref{18_1}) plus the term $\frac{\eta}{\sqrt{3}}G_{K^{*}\bar{K}}\frac{g_{a_1, K^{*}\bar{K}}g_{a_1,\rho\pi}}{M^2_{\mathrm{inv}}(\rho^0\pi^{+})-m^2_{a_1}+im_{a_1}\Gamma_{a_1}}$ with $\rho^0\pi^+\eta$ in the final state. Hence, we see the $\rho$ peak again in the $M(\pi^+\pi^-)$ distribution. Not surprisingly, the $a_1$ term contribution which contains this term in the way we calculate, still shows a peak for the $\rho$. The other thing that we observe is that the interferences are important in order to give the final structures.

The exercise done here, shows the complexity of the current problem, and why a standard fit summing Breit-Wigner structures for resonant excitation, as it is usually done in experimental analysis, must be taken with care. While we get the different important modes reported in~\cite{BESIII:2021aza}, the $a_1$ signal which is the dominant one in~\cite{BESIII:2021aza}, is relevant in our study but not dominant, and the fit fraction of 12.7\% reported in~\cite{BESIII:2021aza} for the $f_0(500)\pi^+$ mode, is absent in our study. However, we could find the origin of the seeming $f_0(500)$ peak in  $M(\pi^+\pi^-)$, because one has two $\pi^+\pi^-$ at the end: one $\pi^+\pi^-$ pair producing the $\rho$ and the other giving rise to this bump.

We would like to finish this discussion with the comment that in the experimental analysis of this reaction $10$ fit fractions and $9$ phases where considered as free parameters. We only have $5$ parameters and  a global normalization. The fact that the resonances studied were all dynamically generated, except for the $\rho$, from pseudoscalar-pseudoscalar or pseudoscalar-vector interaction, and that we could relate to the different weak decay modes the production of these meson meson components, established constraints on the relative production of these resonances. In addition, we also benefitted from having couplings of resonances to the different meson channels, provided by the chiral unitary approach, such that, at the end, we had a significantly smaller freedom to fit the data, in spite of which we could get a fair reproduction of them.

\section{conclusions}
We have made a theoretical study of the $D^+_s\to\pi^+\pi^+\pi^-\eta$ reaction taking into account the final state interaction of pairs of mesons. The pairs are not necessarily those of the final state, because we consider coupled channels, and it is possible to produce some mesons in a first step which lead to the final $\pi^+\pi^+\pi^-\eta$ state making transitions with the strong interaction. The problem is complicated because of the many possible intermediate states but we followed a systematic study in which we looked at the weak decay processes, Cabibbo favored, that stemmed from $D_s\to$ quarks with external and internal emission. Then we allowed for one or two hadronizations of $q\bar{q}$ pairs to obtain mesons. Even then there were many possible channels, but the filter of $G$-parity of these states reduced considerably the number of possible combinations. Also, some of these were shown to be unable to produce the $\pi^+\pi^+\pi^-\eta$ final state after rescattering. This allowed us to have a manageable amount of terms at the end, with only a few parameters correlating them. The transition $t$-matrices needed to go from these selected channels to the final state are all taken from the chiral unitary approach. At the end we have some amplitudes which interfere among them, from where we can get the six mass distributions and compare them with the experiment. The amplitudes that we obtain, through final state interaction generate scalar and axial vector resonances, $a_0(980)$, $f_0(980)$, $a_1(1260)$, and $b_1(1235)$, which are visible in the experimental mass distributions, except for the $b_1(1235)$ which also comes with negligible strength in our study.

We obtain an acceptable fit to the six mass distributions with considerably less freedom in the parameter space than in the experimental analysis, which should be considered as a support for the amplitudes that we obtain using the chiral unitary approach, where the low lying scalar mesons are generated from the pseudoscalar-pseudoscalar interaction and the axial vector resonances from the pseudoscalar-vector interaction. The relevant modes obtained from a ft to the data with a sum of Breit-Wigner structures in the experimental analysis were also found in our study, but we had less strength for the dominant $a_1(1260)$ mode, although we noted that there are large interference of the amplitudes and one must look with caution at the meaning of a fit fraction. An interesting feature is that the $f_0(500)$ was not produced in our approach, but the broad bump seen in the $M(\pi^+\pi^-)$ mass distribution around $500$~MeV came as a consequence of having two $\pi^+$ in the final state, from the `wrong' $\pi^+\pi^-$ pair when the `good' $\pi^+\pi^-$ pair produced the $\rho^0$.

In these mass distributions we could clearly see peaks for the $\rho^0$, $a_0^+$, $a_0^-$, $f_0(980)$ and $a^+_1(1260)$ in reasonable agreement with experiment. But once again we caution about determining fit fractions of these resonances given the large interferences found.

The other finding of our study is that thanks to the explicit consideration of resonances as coming from meson meson scattering in coupled channels, we could reproduce the data starting with mechanisms of external and internal emission and did not need to invoke a weak annihilation mechanism. This is because even if some resonances are not produced in a first step through external or internal emission, some of the coupled channels, different to the state observed at the end, can be produced. The rescattering produces the desired final state, while at the same time generates some resonances.

\section{Acknowledgments}

Jing Song would like to thank Dr. Jun-Xu Lu for useful discussions. This work is partly supported by the Spanish Ministerio de Economia y Competitividad 
(MINECO) and European FEDER funds under Contracts No. PID2020-112777GB-I00, and by Generalitat Valenciana under contract PROMETEO/2020/023.
This project has received funding from the European Union Horizon 2020 research and innovation programme under the program H2020-INFRAIA-2018-1, grant agreement No. 824093 of the STRONG-2020 project. J. S. wishes to acknowledge support from China Scholarship Council. 
The work of A. F. was partially supported by the Generalitat Valenciana and European Social Fund APOSTD-2021-112 , and the Czech Science Foundation, GA\v CR Grant No. 19-19640S.  

\bibliography{refs.bib}
\end{document}